\documentclass[sigplan,screen]{acmart}
\usepackage{amsmath}
\usepackage{algorithm}
\usepackage{booktabs}
\usepackage{float}
\usepackage{multirow}
\usepackage{color}     
\usepackage{algorithmic}  
\usepackage{balance}

\copyrightyear{2026}
\acmYear{2026}
\setcopyright{cc}
\setcctype{by}
\acmConference[ASPLOS '26]{Proceedings of the 31st ACM International Conference on Architectural Support for Programming Languages and Operating Systems, Volume 2}{March 22--26, 2026}{Pittsburgh, PA, USA}
\acmBooktitle{Proceedings of the 31st ACM International Conference on Architectural Support for Programming Languages and Operating Systems, Volume 2 (ASPLOS '26), March 22--26, 2026, Pittsburgh, PA, USA}
\acmPrice{}
\acmDOI{10.1145/3779212.3790128}
\acmISBN{979-8-4007-2359-9/2026/03}

\begin{document}

\title{Architecting Scalable Trapped Ion Quantum Computers using Surface Codes}

\author{Scott Jones}
\orcid{0009-0009-6958-8360}
\affiliation{%
  \department{Department of Computer Science and Technology}
  \institution{University of Cambridge}
  \city{Cambridge}
  \country{United Kingdom}
}
\email{sj665@cam.ac.uk}

\author{Prakash Murali}
\orcid{0000-0003-3378-8589}
\affiliation{%
  \department{Department of Computer Science and Technology}
  \institution{University of Cambridge}
  \city{Cambridge}
  \country{United Kingdom}
}
\email{pm830@cam.ac.uk}

\begin{abstract}
 Trapped ion (TI) qubits are a leading quantum computing platform. Current TI systems have less than 60 qubits, but a modular architecture known as the Quantum Charge-Coupled Device (QCCD) is a promising path to scale up devices. There is a large gap between the error rates of near-term systems ($10^{-3}$ to $10^{-4}$) and the requirements of practical applications (below $10^{-9}$). To bridge this gap, we require Quantum Error Correction (QEC) to build \emph{logical qubits} that are composed of multiple physical qubits. While logical qubits have been demonstrated on TI qubits, these demonstrations are restricted to small codes and systems. There is no clarity on how QCCD systems should be designed to implement practical-scale QEC. This paper studies how surface codes, a standard QEC scheme, can be implemented efficiently on QCCD-based systems. To examine how architectural parameters of a QCCD system can be tuned for surface codes, we develop a near-optimal topology-aware compilation method that outperforms existing QCCD compilers by an average of 3.8X in terms of logical clock speed. We use this compiler to examine how hardware trap capacity, connectivity and electrode wiring choices can be optimised for surface code implementation. In particular, we demonstrate that small traps of two ions are surprisingly ideal from both a performance-optimal and hardware-efficiency standpoint. This result runs counter to prior intuition that larger traps (20-30 ions) would be preferable, and has the potential to inform design choices for upcoming systems.
\end{abstract}



\maketitle

\section{Introduction}
\begin{figure*}[t]
    \centering
       \includegraphics[width=228pt]{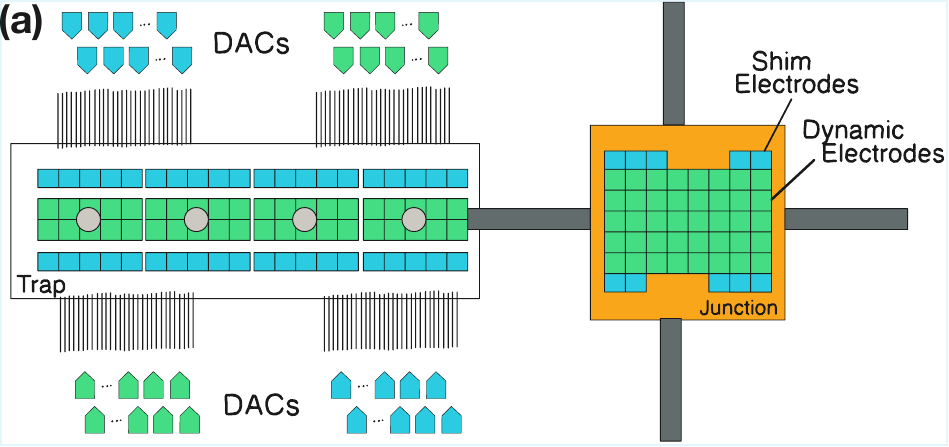}   
\includegraphics[width=266pt]{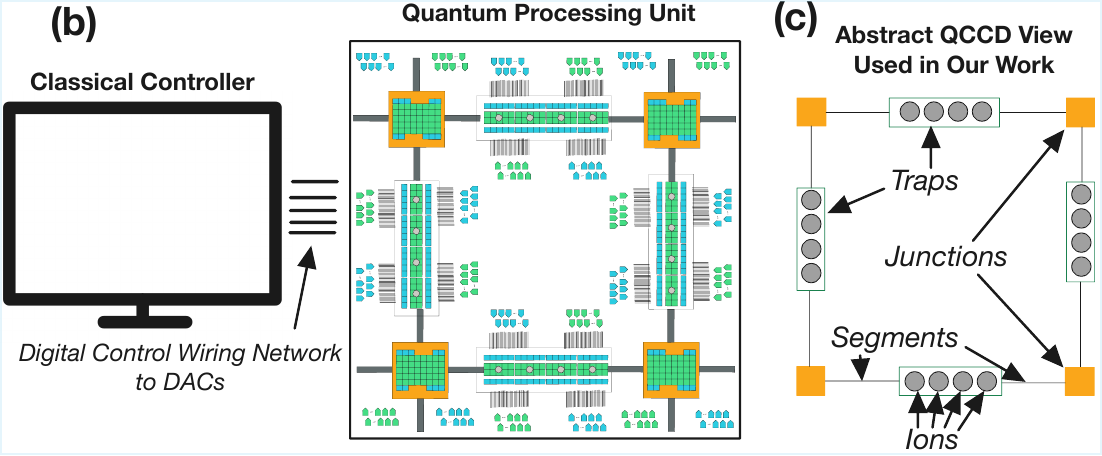} 
\caption{%
            Quantum Charge-Coupled Device (QCCD) system. A detailed view of the QCCD hardware, where ions (grey circles) serve as qubits and are confined within an electromagnetic field known as a trap. 
            \textbf{(a)} The trap is structured with different types of electrodes to position ions, including dynamic electrodes (green) for time-varying signals and shim electrodes (blue) for static potentials. Transport segments (black) and junctions (orange) allow ions to move between traps. 
            \textbf{(b)} The QCCD device is controlled by a classical system interfacing with Digital-to-Analog Converters (DACs), each responsible for individual electrode voltages, enabling precise ion control~\protect\cite{Lekitsch_2017}. 
            \textbf{(c)} We use an abstract QCCD view for this paper.%
        }

    \label{figQCCDHardwareDesign}
\end{figure*}
\begin{figure}[t]
    \centering
    \includegraphics[width=215pt]{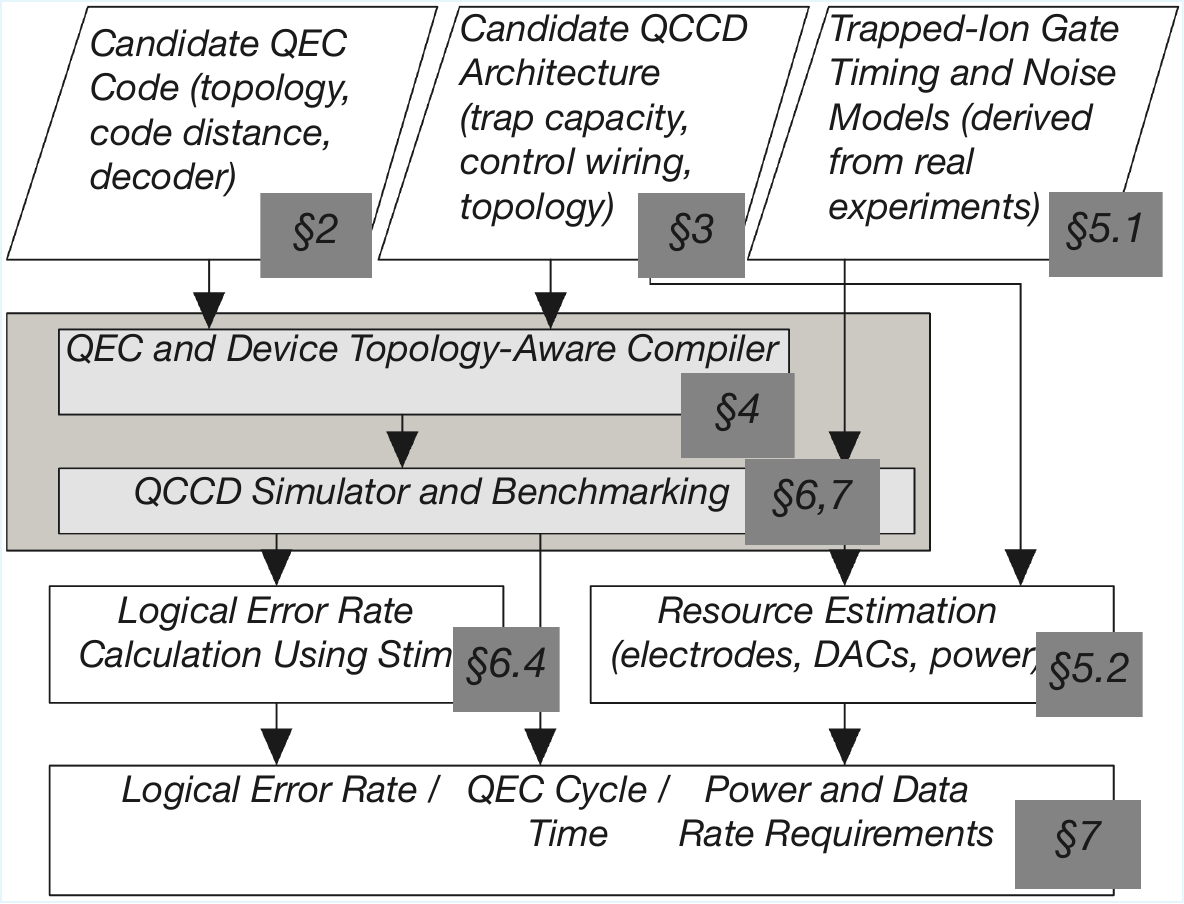}
    \caption{Framework for evaluating the suitability of a candidate QCCD-based TI system for error correction. Taking a candidate architecture and a candidate QEC code as input, the tool flow computes error correction metrics such as logical error rate, QEC round time and power dissipation requirements by using a QEC and device topology-aware compiler, QCCD simulator, and realistic models for performance and resource estimation.}
    \label{figPaperOverview}
\end{figure}

\par Trapped ions (TI) qubits are an important platform for realising scalable quantum computers. Over a hundred academic groups are pursuing this technology \cite{QIPGroups}, and production systems have been demonstrated by IonQ, Quantinuum, Oxford Ionics and other vendors~\cite{ionq, quantinuum, oxfordionics}. Small TI systems use a monolithic architecture where all qubits are housed in the same physical trap. This design is not scalable due to control challenges and poor gate fidelities (quality of gate operations), especially beyond 30 qubits \cite{murali2020architectingnoisyintermediatescaletrapped,PhysRevA.98.032318}. Instead, a modular design where ions are distributed across many small traps is seen as a path towards scalable systems. This architecture, termed the Quantum Charge-Coupled Device (QCCD) was first proposed in 2002~\cite{Kielpinski2002} and has been demonstrated in practice by Quantinuum~\cite{H2Racetrack}. Figure \ref{figQCCDHardwareDesign} shows an example QCCD system with four traps.

\par To achieve a practical quantum advantage over classical computing, we require \(\approx 100-1000\) algorithmic qubits with an error rate of at least $10^{-9}$ \cite{suppressing_quantum_errors_by_scaling_surface_code}, which is well beyond the limits of all known qubit technologies \cite{IBMQuantumRoadmap, GoogleQuantumAIRoadmap}. Therefore, we require quantum error correction (QEC). Similarly to classical error correction, where bits are redundantly encoded, QEC encodes a \emph{logical qubit} across multiple physical qubits, detecting and correcting errors. The surface code \cite{Fowler2018LowOQ} is among the most promising candidates for QEC codes due to its compatibility with planar architectures. In this paper, we study how surface code-based logical qubits can be efficiently implemented on QCCD hardware. Although our work focuses on the surface code, our techniques and framework are more broadly applicable.

For two reasons, implementing scalable surface code logical qubits in a QCCD architecture is non-trivial. First, QCCD systems offer a rich architectural design space with a range of trap capacity (number of ions per trap), communication topology (wiring between traps) and control wiring (hardware responsible for orchestrating ion movement) choices. The performance of the surface code logical qubit and its logical error rate depend heavily on the underlying device architecture. \textit{How should device architects navigate these choices for logical qubit implementation?} Second, the performance of the surface code also depends on its mapping to the hardware and the routing techniques that are used to orchestrate the movement of ions. \textit{How can we optimise these mappings across various architectures and surface code parameters?} Previous work and industry roadmaps either focus on noisy intermediate-scale quantum (NISQ) workloads~\cite{murali2020architectingnoisyintermediatescaletrapped} or use manual mappings~\cite{Lekitsch_2017} or only pick out a few architectural choices without rigorous architectural exploration~\cite{valentini2024demonstrationtwodimensionalconnectivityscalable}.

Our work performs the first systematic design space exploration for logical qubit implementation on QCCD devices. 
We require an efficient compilation of surface code parity-check circuits (Figure~\ref{fig:RotatedSurafecCode4WithQC}) onto diverse QCCD architectures to enable architecture evaluation. 
Only a few compilers \cite{Sivarajah_2020, murali2020architectingnoisyintermediatescaletrapped, muzzletheshuttle} support QCCD, but they are designed for NISQ applications on small QCCD hardware \cite{Moses_2023}. 
We developed a novel QEC and device topology-aware mapping scheme that exploits the parallelism and structure inherent in the parity check operations in the surface code to find good mapping solutions.
Our compiler maps logical qubits to hardware and then implements logical qubit instructions using low-level QCCD primitives while adhering to QCCD hardware constraints. 
Using this compiler, we develop a toolflow for design space exploration, shown in Figure~\ref{figPaperOverview}. 
This toolflow accepts a QEC code and QCCD architecture as input and then arrives at an efficient mapping, which is used alongside architectural models and logical qubit simulations to determine metrics such as cycle time, logical error rate and data rate. 
We use the tool to sweep the architectural design space and select optimal designs. \textbf{Our contributions are as follows:}

\begin{itemize}
 \item We identify important architectural parameters for the implementation of surface codes in QCCD systems. Unlike previous works\cite{murali2020architectingnoisyintermediatescaletrapped}, we identify that a trap capacity of two ions is surprisingly ideal even though it maximises communication operations. When paired with grid connectivity and direct wiring of electrodes to DACs, we can achieve near-optimal cycle times and low logical error rates across both small and large surface code implementations, compared to higher trap capacity configurations. 
\item Comparing WISE\cite{1000qubits}, a state-of-the-art wiring method with the standard QCCD architecture, we identify a power vs. cycle time scaling bottleneck. Existing wiring methods either offer high power with fast logical clock speeds or low power but very slow speeds. For near-term demonstrations, these techniques are sufficient. However, to scale up to hundreds of logical qubits, we require a fundamental re-design of the wiring architecture considering power consumption as part of hardware-software co-design. 
\item Our QEC and device topology-aware compiler offers near-optimal QEC round times, outperforming existing compilers by 3.8X \cite{muzzletheshuttle, murali2020architectingnoisyintermediatescaletrapped}. Unlike existing QEC compilers for QCCD systems, our compiler can handle large surface code implementations and scale to large trap capacities.
\end{itemize}

\section{Background}\label{subsecTopoCodesBackground}

\par \textbf{Trapped ions:} In a TI system, the ions act as qubits.  For example, a popular choice is a Calcium ion. To hold ions in place, an electromagnetic field is used. This field is generated using DC electrodes.  As a result of this control mechanism, the ions are arranged as a linear chain. Single-qubit gates are implemented using a laser to excite a specific ion, while two-qubit gates involve multiple lasers that excite the internal states and shared vibrational motion of ions within the same trap.

\begin{figure}[t]
    \centering
    \includegraphics[width=238pt]{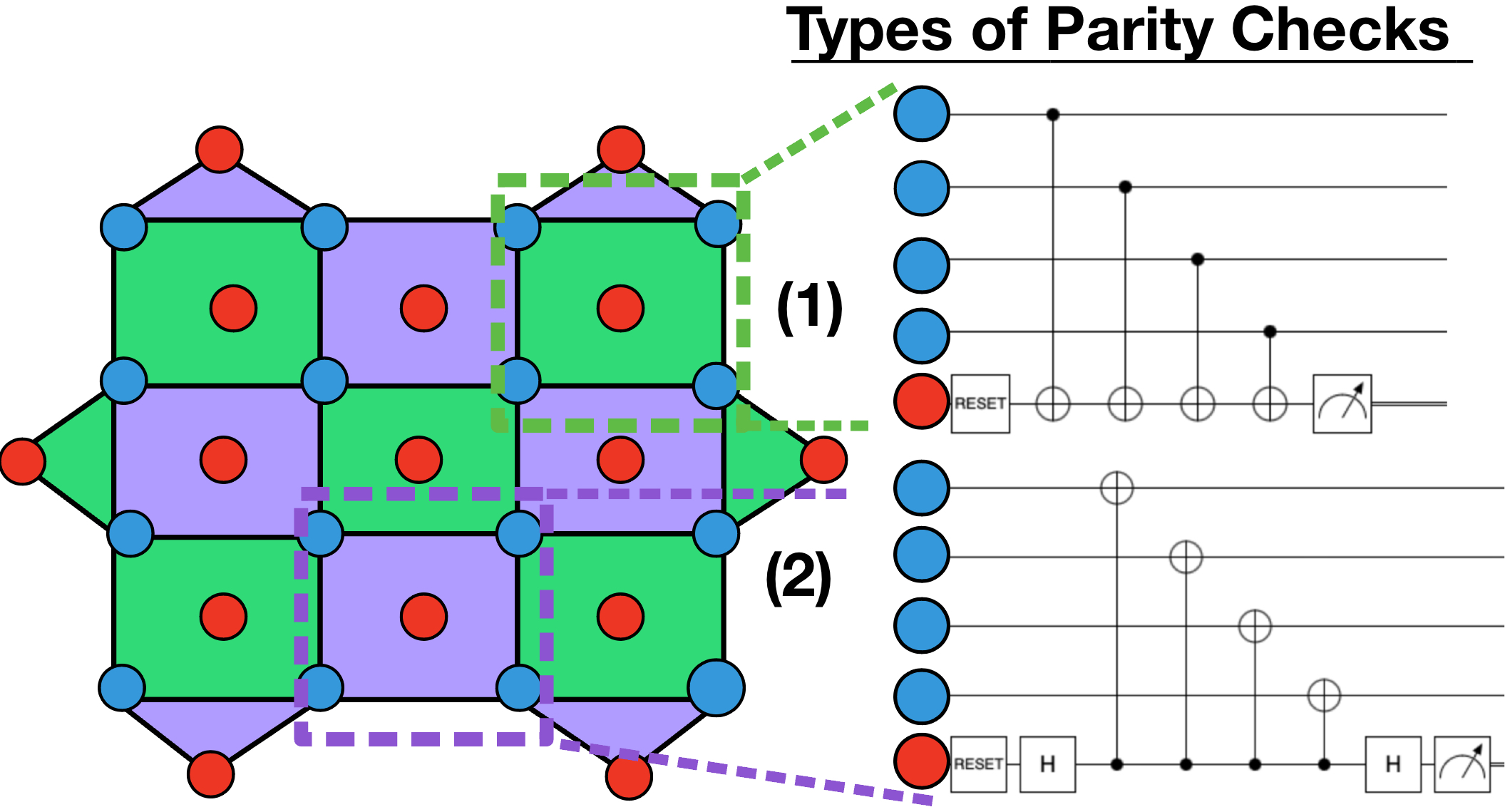}
    \caption{The topology of the distance four surface code. The blue circles represent physical data qubits, and the red circles represent physical ancilla qubits. Data qubits form the vertices of the cells that make up the shaded surface, and there is exactly one ancilla qubit in the centre of every cell. The cells are shaded purple or green to disambiguate the two types of parity checks, with each type of circuit given on the right.  }
    \label{fig:RotatedSurafecCode4WithQC}
\end{figure}
\par \textbf{Surface codes:} Figure~\ref{fig:RotatedSurafecCode4WithQC} illustrates a surface code qubit. Surface codes are a family of QEC codes that encode a logical qubit into a planar \(d \times d\) grid of physical qubits, called data qubits, where \(d\) is the \emph{code distance}. QEC is effective only when the physical error rate of the qubits in the hardware is below the \emph{code threshold}\label{def:code_threshold}. Below the threshold, a larger code distance offers exponentially lower logical error rates at the expense of more physical qubits per logical qubit (scaling as $O(d^2)$).

We focus on the surface code due to its high code threshold and ease of hardware implementation. This is because most quantum circuits for the surface code are a regular set of parity checks, where every ancilla (red) qubit is initialised, then the ancilla has a two-qubit entanglement gate with only each of its 4 neighbouring data (blue) qubits, and finally the ancilla qubit is measured (shown on the right of Figure~\ref{fig:RotatedSurafecCode4WithQC}). It is a well-accepted choice for TI systems \cite{Lekitsch_2017}.

\par \textbf{Primitive QCCD Operations:} \label{def:QCCD_toolbox}

We use a set of primitive operations that provide the quantum gates necessary to maintain a logical qubit \cite{transversality_lattice_surgery}. The entangling gate is a two-qubit Mølmer–Sørenson (MS) gate (t1); the implementation details are not relevant for this paper. Single-qubit gates are rotations around the x, y, and z axis on a single isolated ion (t2-t4). In addition, there are (t5) measurements of trapped-ion qubits and (t6) qubit reset. QCCD movement techniques include (t7) shuttling (moving) an ion across a transport segment connecting one trap or junction to another, (t8) splitting (moving an ion from a trap into a segment) and (t9) merging (moving an ion from a segment into a trap). An ion must be at the end of a trap in order to split (t8), which can be done by swapping ions within a trap (via 3 two-qubit gates (t1)).  The final primitives are (t10) junction crossing entry and (t11) exit, whereby ions move across junctions that connect different segments. We assume that only a single ion could reside in a junction and that only a single ion could reside in a single segment at any moment \cite{Burton_2023,PhysRevLett.109.080501, PhysRevLett.109.080502}.

\section{QCCD Logical Qubit Design Trade-offs} \label{sec:hardware_tradeoffs}

\subsection{Trap Capacity} \label{subsec:trap_capacity}
\par A key architectural choice for QCCD systems is trap capacity, defined as the maximum number of qubits per trap. For example, Figure \ref{figQCCDHardwareDesign} shows a trap with capacity 4. There are three aspects to the choice of trap capacity. First, with high capacity, inter-ion spacing reduces and makes it difficult to address individual ions in the trap with laser controllers \cite{murali2020architectingnoisyintermediatescaletrapped}, leading to poor gate fidelity. Second, with high capacity, the need for communication operations is reduced. This can improve overall circuit fidelity due to a shorter depth and the reduction in the number of noisy operations. Third, in typical trapped-ion QC implementations, the gates within the same trap are executed serially. Although parallel two-qubit gates have been demonstrated \cite{Figgatt_2019}, these gate times are 6X worse than the sequential gate times we assume and the gates have been challenging to realise beyond small scales \cite{murali2020architectingnoisyintermediatescaletrapped}. To our knowledge, current QCCD platforms (IonQ, Quantinuum) do not offer parallel two-qubit gates within a trap for this reason \cite{Chen_2024}. Therefore, QCCD systems with multiple small traps can execute more gates in parallel, reducing the overall execution time.

\par While prior works have explored the choice of trap capacity for NISQ workloads \cite{murali2020architectingnoisyintermediatescaletrapped}, \textbf{the optimal trap capacity for logical qubit implementation with QCCD systems is unknown}. For surface code logical qubits, there are intuitive choices for this parameter. For example, each qubit can be mapped to a separate trap. This offers the highest possible two-qubit gate fidelity at the expense of many communication operations. Similarly, non-adjacent parity checks, shown on the right of Figure~\ref{fig:RotatedSurafecCode4WithQC}, can each be mapped to a trap with capacity 5. This reduces communication compared to the former case. As an extreme choice, the entire logical qubit in Figure \ref{fig:RotatedSurafecCode4WithQC} can be mapped to a single trap with capacity 31 (IonQ's systems adopt this approach \cite{Chen_2024}). As discussed, this serialises operations and kills the inherent parallelism available.

\subsection{Communication Topology} \label{subsec:communication_topo}

To determine the optimal trap capacity, it is crucial to consider the communication topology of the QCCD device. The choice of topology determines the number of ion transport operations (t7-t11 §\ref{def:QCCD_toolbox}) that will be required. Ions have all-to-all connectivity within a trap, while ions in different traps are connected by shuttling paths, which are implemented using segments and junctions in hardware (Figure~\ref{figQCCDHardwareDesign}). Unlike general NISQ workloads with widely varying communication requirements \cite{murali2020architectingnoisyintermediatescaletrapped}, surface code parity-check circuits have a regular local structure. As a result, ion movement operations can remain local if the communication topology between ions preserves the structure of the surface code. For example, a grid topology, where traps are interconnected by a grid network of shuttling paths and junctions (Figure~\ref{figQCCDHardwareDesign}), closely aligns with the structure of the surface code when trap capacity is minimal \cite{Lekitsch_2017}. However, \textbf{the performance of the grid topology is unclear when large trap capacities are used}. Further, we consider two more topologies as optimistic and pessimistic cases: an all-to-all switch topology where traps are connected using an n-way junction and a linear topology where all traps are connected to their nearest neighbour on a line. The optimistic case loosely resembles the MUSIQC architecture proposed for trapped ions \cite{PhysRevA.89.022317}, and the pessimistic case resembles the architecture of Quantinuum's current H-series devices \cite{H2Racetrack}.

\subsection{Control System Wiring Choices}
\label{subsec:WISE_network_intro}

\begin{figure}[!htbp]
    \centering
    \includegraphics[width=245pt]{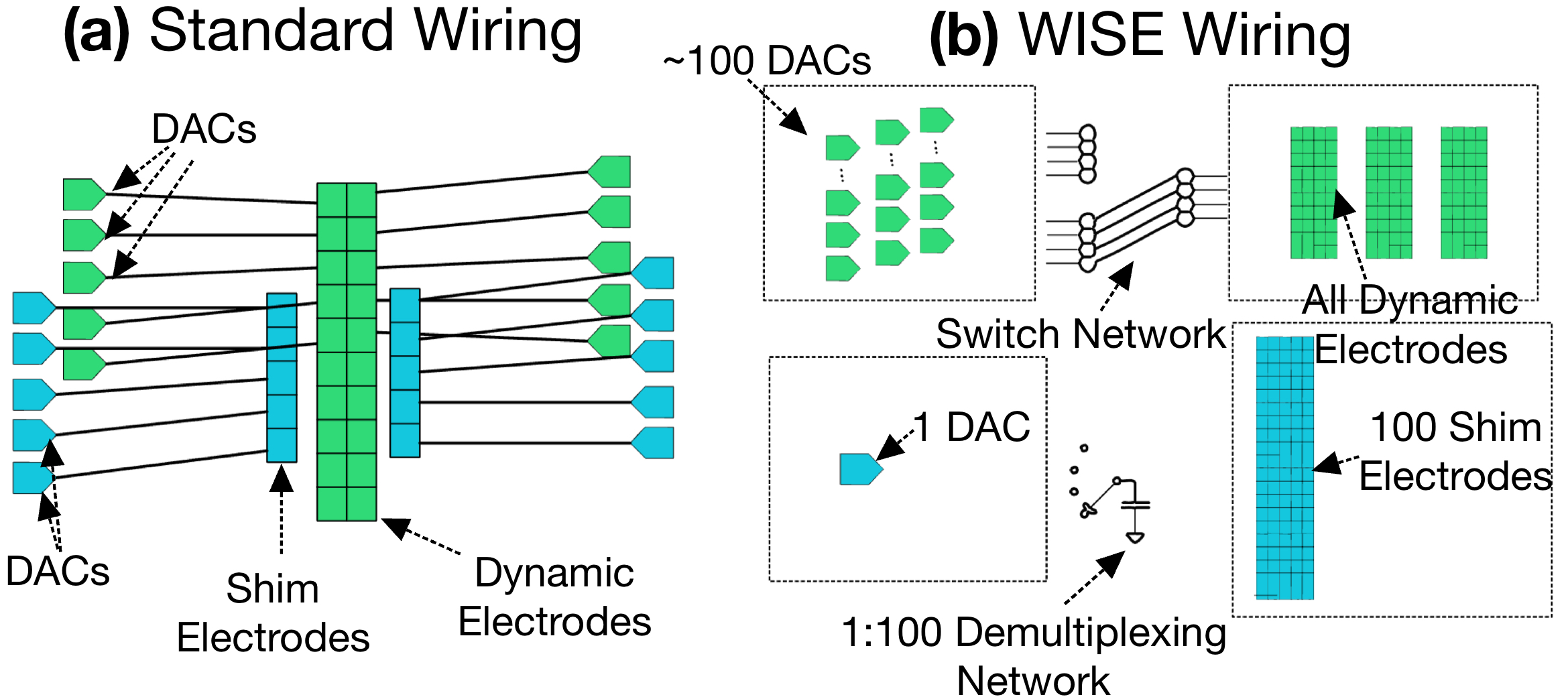}
    \caption{\textbf{(a)} Each electrode is connected to a dedicated DAC in the standard architecture, resulting in a direct but highly resource-intensive wiring scheme. \textbf{(b)} The WISE architecture integrates an ion trap with a switch-based demultiplexing network, significantly altering the scaling of control electronics. All dynamic electrodes (green) are controlled with \(\approx 100\) DACs irrespective of system size by using a switch network, but this comes at the cost that only primitive QCCD operations of the same type (t1-t11 §\ref{def:QCCD_toolbox}) can execute simultaneously. One DAC can set \(\approx 100\) shim electrodes (blue).}
    \label{fig:StandardvsWISEwiring}
\end{figure}

\par Another key aspect of scaling trapped-ion QCCD systems for fault-tolerant quantum computation is managing control electronics. \textit{How should electrodes (used to position and move ions) be wired to the digital-to-analog converters (DACs) which control trap voltages?} Traditional QCCD architectures employ one DAC per electrode (Figure~\ref{fig:StandardvsWISEwiring}). Each ion qubit requires tens of electrodes, and therefore, the number of control signals needed for implementing large surface code qubits becomes impractical. For instance, a distance 7 surface code (with 49 physical qubits) requires 5500 DACs per logical qubit, which is equivalent to \(\approx 275\)GBit/s controller-to-QPU bandwidth (§\ref{subsecresourcegstimation}). 

\par One leading alternative is the Wiring using Integrated Switching Electronics (WISE) architecture \cite{1000qubits}, which integrates a \emph{switch-based demultiplexing network} (bottom of Figure~\ref{fig:StandardvsWISEwiring}). By sharing a smaller set of DACs across many electrodes, WISE scales more favourably regarding control complexity and power consumption. However, this benefit comes with a critical trade-off: only one type (t7-t11 §\ref{def:QCCD_toolbox}) of ion movement primitive can co-occur, restricting parallelism in ion routing.

Given a QCCD architecture, the logical error rate of the surface code implementation and its cycle time are the two most important metrics that guide system design. Therefore, we ask \textit{``What is the optimal trap capacity to achieve practical logical error rates for realistic surface code distances and logical clock speeds? Does the grid topology offer good code performance across a range of trap capacities? What is the best current wiring method? Does the reduced hardware overhead in WISE justify the longer logical clock speeds, or is the standard scheme more practical for achieving logical error rates less than \(10^{-9}\)?''}

\section{Topology-Aware QEC-to-QCCD Compiler} \label{sec:logicalToHardwareQubits}
\begin{figure*}[!htbp]
    \centering
    \includegraphics[width=350pt]{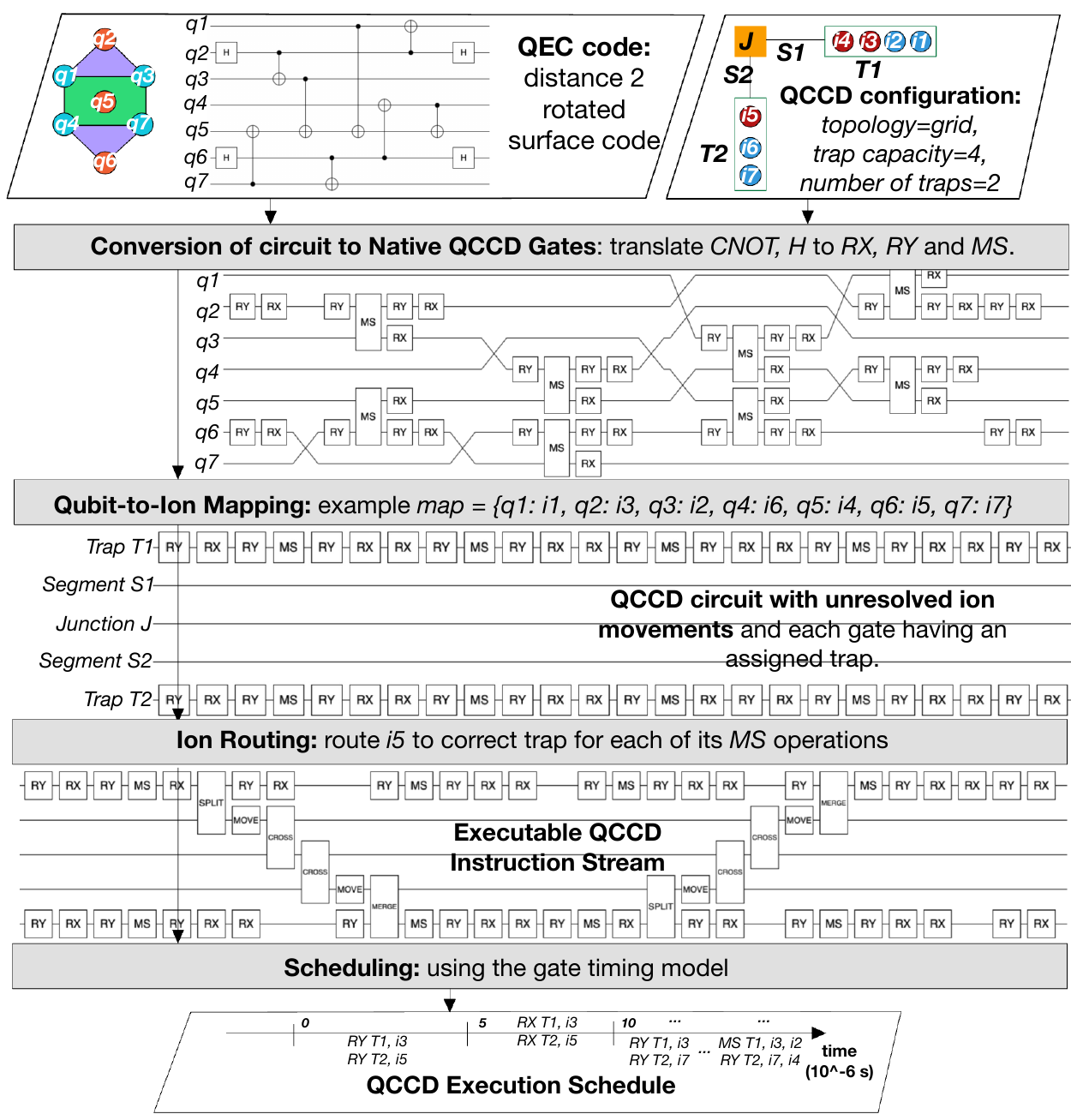}
    \caption{QCCD compilation flow: from a distance 2 surface code (syndrome extraction) circuit (top-left) and QCCD device configuration (top-right) to a scheduled, executable QCCD program. Steps include translation to native gates, qubit-to-ion mapping, ion routing using the movement primitives from the QCCD toolbox (§\ref{subsecTopoCodesBackground}), and scheduling using the operation timings in Table~\ref{tab:operations}.}
    \label{fig:compilationFlow}
\end{figure*}

We require a resource-efficient mapping of the surface code onto QCCD systems with different architectures to answer the design questions. Although several tool flows have been developed to map NISQ workloads on QCCD hardware, they incur large communication overheads and do not scale to high code distances. In this section, we develop a surface code compiler shown in Figure~\ref{fig:compilationFlow}. The compiler takes a surface code and QCCD configuration as inputs. Then, the surface code parity-check circuit is translated into native gates (§\ref{sec:qubitGatesToPrims}). Each surface code qubit is then assigned to a physical qubit in the hardware (§\ref{secQubitToIon}) and reconfiguration operations are inserted into the circuit to ensure that all two-qubit gates occur within the same trap (§\ref{subsec:ion_routing_algo}). Finally, the circuit is converted into an execution schedule (§\ref{sec:scheduling}). Our goal is to minimise the main physical drivers of logical error rate (detailed in Section~\ref{sec:modelling}) under fixed hardware constraints. To ensure scalability at large code distances, we cannot directly minimise the complex error objective function during compilation. Instead, we optimise two proxy metrics: overall circuit runtime (primary) and number of routing operations (secondary) which serve as effective heuristics for minimising the dominant drivers of error, idling noise and ion heating, respectively.

\subsection{Mapping QEC Instructions to QCCD Instructions} \label{sec:qubitGatesToPrims}
Surface code parity-check circuits are expressed in terms of Hadamard, CNOT and measurement operations. These operations are converted into sequences of MS operations (t1) and single-qubit rotations (t2-t4) from the QCCD toolbox (§\ref{def:QCCD_toolbox}) using known gate identities \cite{figgatt2018building}. This is a straightforward intermediate-representation transformation. 

\subsection{Mapping Qubits to Ions} \label{secQubitToIon}
\begin{figure}[!htbp]
    \centering
    \includegraphics[width=245pt]{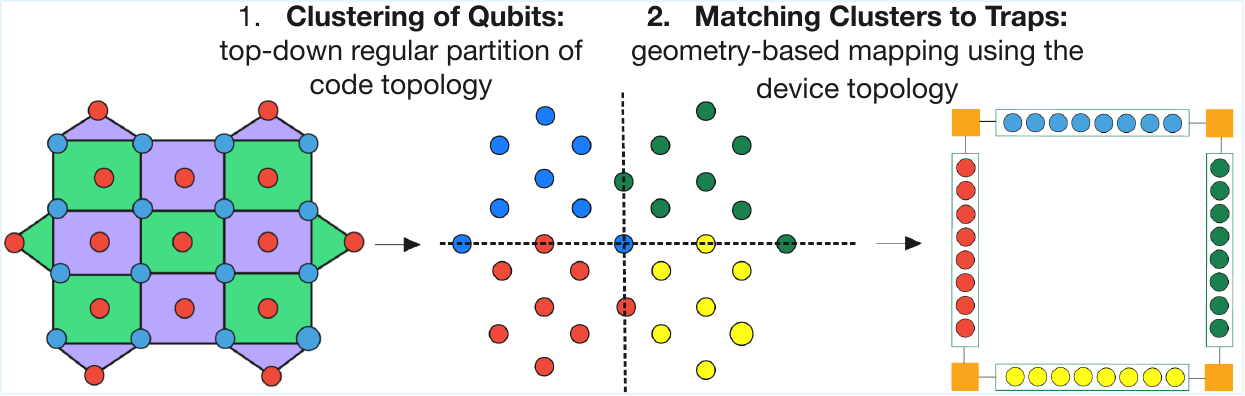}
    \caption{Mapping qubits to ions. Given a distance 4 surface code (left) and a QCCD device with trap capacity 9, we first partition into \(ceil(N_{qubits} / (capacity-1)) = ceil(31 / 8 ) = 4\) clusters of qubits by top-down regular partitioning of the code topology (recursively bisecting the code's qubit layout). The surface code’s regular structure means neighbouring qubits that share entanglement operations are likely grouped into the same cluster, reducing inter-trap communication. Clusters are then mapped to traps using a geometry-based matching that preserves local neighbourhoods, ensuring that qubits in different clusters but adjacent in the code are placed in neighbouring traps, minimising ion movement overhead.}
    \label{figQubitToIonMap}
\end{figure}
The second pass in Figure~\ref{fig:compilationFlow} assigns each qubit in the surface code to a unique physical qubit in the hardware. To determine the mapping, 1) we cluster the qubits into balanced partitions and 2) map the clusters to traps using a matching algorithm. Since there is all-to-all connectivity within a trap, the mapping of individual qubits in the cluster to trap qubits makes almost no difference in the overall execution schedule. 

\par Mappings that fill traps well below capacity will increase the number of ion movements. Similarly, filling traps to maximum capacity is generally inefficient \label{exp:max_capacity_inefficient} , as incoming ion movement would require displacing another ion. We adopt a design where the traps are filled to \(capacity-1\), leaving one ion position free for communication.

\textbf{1. Clustering of qubits:} To partition qubits (into clusters of size \(capacity-1\)), we can solve a balanced graph partitioning problem. Given a graph \( G = (V, E) \), where nodes \( V \) represent qubits and edges \( E \) represent pairs of data and ancilla qubits undergoing entanglement operations, with edge weight proportional to the order of operations in the circuit (early operations have high weight), the objective is to divide \( V \) into equal-sized clusters \( C_1, \ldots, C_k \) such that the total weight of cut edges is minimised. Here, \( k \) will equal the number of traps used by the logical qubit in the QCCD hardware. The number of ion movement operations is minimised by minimising the number of high-priority entanglement operations cut. Note that balancing improves execution time due to fewer ion reconfigurations, which decreases the logical error rate when qubits are noisy. Balancing does not affect correctness: all partitions result in correct sequences of operations for the surface codes if the underlying qubits are perfect.

\par In general, the balanced graph partitioning problem is NP-complete \cite{garey1979computers} and has no finite factor polynomial-time approximation when partitions must be exactly equal \cite{balanced_graph_partitioning}. Therefore, other compilers \cite{Sivarajah_2020, muzzletheshuttle, murali2020architectingnoisyintermediatescaletrapped} that are designed for general quantum circuits are not able to efficiently cluster qubits for large code distances. Whereas, for regular grid-like graphs typical of surface codes, our compiler can approximate a balanced partition well. We use a top-down regular partitioning of the surface code topology, as depicted in Figure~\ref{figQubitToIonMap}. This minimises ancilla movement between traps because qubit neighbourhoods are preserved in the map, and the surface code only contains entanglement operations between neighbouring qubits. Minor imbalances (by 1–2 qubits) can occur due to code boundary effects.

\textbf{2. Mapping of clusters to traps:} Clusters are then mapped to traps by solving a minimum edge-weight, maximum cardinality matching problem,  which results in a geometry-based mapping, as depicted in Figure~\ref{figQubitToIonMap}, ensuring that the neighbours of each qubit that belong to different clusters still reside in neighbouring traps. 

\par In the matching problem, the edges between clusters and traps are weighted by the distance between the centre of qubit clusters in the code topology and the trap positions in the hardware topology. The problem is solved by considering all subsets of traps with cardinality equal to the number of clusters, where, for each subset, we use the Hungarian algorithm \cite{Kuhn1955Hungarian} to compute the minimum perfect matching in polynomial time, and the subset with the lowest total cost is selected. For general quantum circuits, the search space can be reduced to an exponential number of trap subsets by considering only contiguous subsets (no holes) whose centres lie near the centre of all traps in the hardware. To achieve polynomial-time compilation, we further prune subsets using patterns in the boundary of the surface code topology. The compiler generalises to other scalable QEC codes, since they are expected to adhere to grid-like structures compatible with the grid QCCD communication topology, making the compiler suitable for expected real-world applications.

\subsection{Ion Routing Algorithm} \label{subsec:ion_routing_algo}
\par To be able to execute an entanglement operation between ions located in different traps, the compiler must determine the appropriate sequence of ion movement operations to ensure that both ions co-exist in the same trap. The physical state of the QCCD architecture during ion routing is modelled as a directed graph where nodes, representing traps and junctions, track the position of each ion, while edges in the graph track the sequence of movement primitives required to transfer an ion between nodes. For each ancilla qubit, the shortest path from the source to the destination trap is determined in the directed graph, and then edge labels along this route are concatenated for the sequence of primitives needed to move the qubit.

\par The ion routing algorithm computes a shortest path for each ancilla qubit to reach its corresponding data qubit's trap while satisfying QCCD hardware constraints:

\begin{itemize}
    \item \textbf{Trap capacity:} Each trap has a fixed maximum ion count at any time \cite{murali2020architectingnoisyintermediatescaletrapped,PhysRevA.98.032318}.
    \item \textbf{Junction exclusivity:} Only one ion may occupy a junction at any time \cite{Burton_2023}.
    \item \textbf{Segment exclusivity:} Only one ion may occupy a shuttling segment at any time \cite{PhysRevLett.109.080501,PhysRevLett.109.080502}.
\end{itemize}

\par
\begin{figure}[!htbp]
    \centering
    \includegraphics[width=234pt]{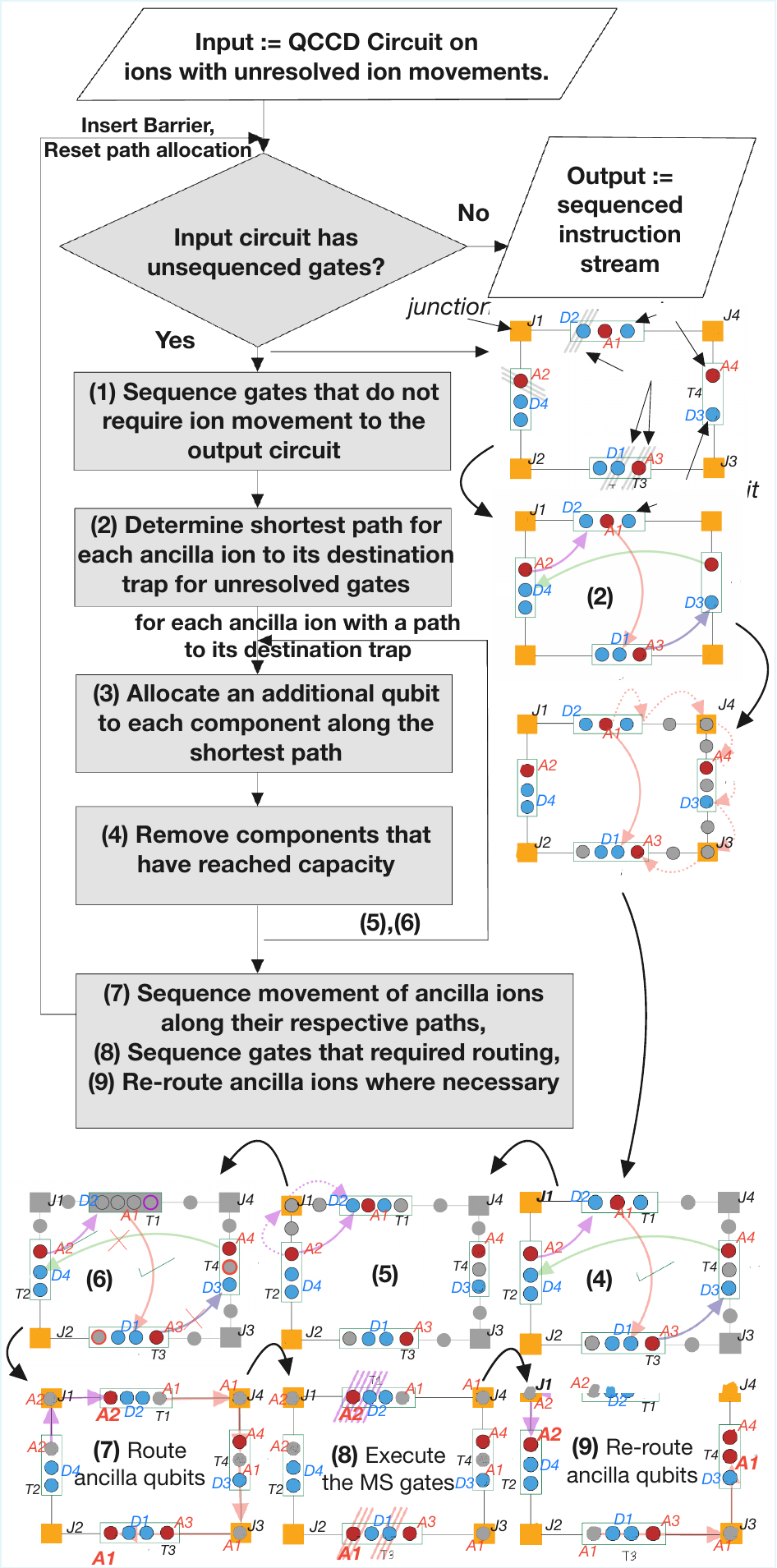}
    \caption{Ion Routing. This example tracks the evolution of ion positions (circles) during a single routing pass. (1) Gates that are not blocked by other unsequenced gates and do not need routing are sequenced. (2) The destination traps for each ancilla qubit are determined based on their next operation. Routing paths are allocated sequentially to ancilla qubits, prioritising those needed earlier in the input gate sequence. (3) Finds a path for ancilla A1, with each component along the path (except the source) being allocated a qubit. (4) Grey components (J4 and J3) have reached capacity, so they are removed (along with T4) from the QCCD graph. (5-6) repeat process (3-4) for ancilla A2. In (6), neither A3 nor A4 can be routed, so (7) proceeds to insert the movement primitives of A1 and A2 along their allocated paths. (8) Sequences the gates that require routing for A1 and A2. (9) Re-routes A1 and A2 to T4 and T2, respectively, to ensure traps are at least 1 below capacity.}
    \label{fig:routing_algo}
\end{figure}

 Once the QCCD graph is constructed, the routing algorithm processes the sequence in multiple passes, sequencing the primitive operations into the output instruction stream until none remain. At the start and end of each pass, each trap is at most one ion below its capacity, and no junction nor segment contains an ion. These invariants ensure that the trap capacity constraint is met during execution.  Each pass of the algorithm is described in Figure~\ref{fig:routing_algo}.

\subsection{Scheduling} \label{sec:scheduling}
\par During routing, a happens-before relation is constructed between operations, producing a logically correct instruction stream. Figure 7 walks through a concrete example where we apply one pass on a small QCCD instance, with each subfigure showing the state after each step. The scheduler then uses the operation times from Table~\ref{tab:operations} to assign physical timestamps to operations in the stream, following a dependency-preserving transformation. The scheduler uses a standard list scheduling approach for precedence- and resource-constrained operations~\cite{graham1966bounds,pinedo_scheduling}, where  ready operations are prioritised by time-weighted critical path length to reduce total circuit runtime~\cite{hu1961parallel}.

\section{Modelling Logical Qubits in QCCD}\label{sec:modelling}
\begin{table}[t]
    \centering
    \footnotesize
    \begin{tabular}{|l|c|c|}
        \hline
        Operation & Duration & Infidelity \\ \midrule \hline
        (t1) Two-qubit MS gate          & $40 \mu s$ & (Refer to \ref{qubit_gate_fidelity_model})   \\ \hline
        (t2-t4) Ion Rotation             & $5 \mu s$    & (Refer to \ref{qubit_gate_fidelity_model})  \\ \hline
        (t5) Measurement                 & $400 \mu s$ & $1 \times 10^{-3}$   \\ \hline
        (t6) Qubit reset                & $50 \mu s$    & $5 \times 10^{-3}$  \\ \hline
        (t7) Ion shuttling              & $5 \mu s$   & $\bar{n} < 0.1 $      \\ \hline
        (t8-t9) Ion split and merge        & $80 \mu s $  & $\bar{n} < 6 $      \\ \hline
        (t10-t11) Junction entry/exit & $100 \mu s$  & $\bar{n} < 3 $  \\  
        \hline
    \end{tabular}
    \caption{Operating parameters for QCCD systems derived from ~\protect\cite{transversality_lattice_surgery}. The reconfiguration steps (t7–t11) do not directly cause gate infidelity; however, they introduce idling noise and increase subsequent gate error rates due to heating, quantified using the mean vibrational energy \(\bar{n}\). For our analysis, we pessimistically use the upper bound values.}
    \label{tab:operations}
\end{table}
This section uses the compiled parity-check circuit to determine its hardware performance, logical error rate, and physical resource requirement. While the focus here is on device modelling, this step is essential for the architectural evaluation in Section~\ref{sec:results}. It serves as our evaluation tool, translating the compiler's execution schedule into concrete metrics and capturing non-trivial physical effects that drive the trade-offs between different QCCD designs.

\subsection{Performance and Noise Models} \label{subsec:qccd_noise_model}

\par To determine the performance of a QCCD system for surface codes, we use a realistic performance and noise model for each primitive operation based on prior work, shown in Table~\ref{tab:operations}. The runtime of the compiled circuit is calculated using the schedule of operations and the duration of each operation in Table~\ref{tab:operations}. 

\par Determining the logical error rate of the code is more involved and requires a noise simulation. We use Stim simulations for this purpose \cite{gidney2021stim}. The input to Stim is a hardware noise model, which in our case is an error-model taken from experimentally grounded modelling of dominant noise sources in trapped-ion hardware, as described in \cite{transversality_lattice_surgery}. In addition, the model has been modified to account for the dependence of qubit gate error rates on the vibrational energy of ions, the number of ions, and the gate duration, as outlined in \cite{murali2020architectingnoisyintermediatescaletrapped}. 

\par \label{def:qccd_error_model} In QEC, physical errors can be decomposed into one of three Pauli channels: X for bit flip, Z for phase flip, or \( XZ = Y\) for bit and phase flip. Our error model incorporates five independent noise parameters to account for the leading experimental imperfections, with different stochastic channels of Pauli errors for various operations:
\begin{enumerate}
    \item \textbf{Dephasing \(e_1\):}\label{depphasing_errors} During ion chain-reconfiguration operations or when qubits are idle, Pauli \(Z\) errors occur with a probability \(p(e_1)\) to account for collective qubit dephasing:
    \[
    p(e_1) = \frac{1 - \exp(-t / T_2)}{2},
    \]
    where \(t\) is the duration of the operation and \(T_2 = 2.2\) seconds is the coherence time for the trapped-ion qubit, obtained from real experiments that demonstrated its accuracy \cite{transversality_lattice_surgery}.
    \item \textbf{Depolarising errors after single-qubit gates \(e_2\):} After single-qubit rotations, Pauli errors (\(X\), \(Y\), or \(Z\)) occur with equal probability \(p(e_2)/3\).
    \item \textbf{Depolarising errors after two-qubit gates \(e_3\):} two-qubit Pauli errors (e.g. two bit flips (XX) or bit flip and phase flip (XZ)) occur with equal probability \(p(e_3)/15\).
    \item \textbf{Imperfect qubit reset \(e_4\):} This is modelled as bit-flip (\(X\)) error occurring after qubit reset to the \(\vert 0 \rangle\) state, with probability \(p(e_4) = 5 \times 10^{-3}\).
    \item \textbf{Imperfect qubit measurement \(e_5\):} This is modelled as bit-flip (\(X\)) errors occurring during measurement with probability \(p(e_5) = 1 \times 10^{-3}\).
\end{enumerate}
\label{qubit_gate_fidelity_model}Errors from ion movement are incorporated into the fidelity model, obtained from \cite{murali2020architectingnoisyintermediatescaletrapped}, which influences the probabilities of errors \(e_2\) and \(e_3\). The fidelity of the qubit gate is influenced by two primary factors: background heating from the trap's electromagnetic field and thermal motion from higher vibrational energy of the ion chain. The fidelity \(p(e_2), p(e_3)\) is expressed as:
\[
p(e_2), p(e_3) = 1 - \Gamma \tau - A (2\bar{n} + 1),
\]
Where \(\Gamma\) is the background heating rate of the trap, \(\tau\) is the gate duration,  \(A \propto \frac{\ln(N)}{N}\) is a scaling factor representing thermal instabilities of the laser beams perpendicular to the ion chain, where \(N\) is the number of ions in the chain, and \(\bar{n}\) is the vibrational energy of the ion chain, quantified in motional quanta (average energy state occupied). The term \(\Gamma \tau\) accounts for fidelity loss due to background heating, which increases with the gate duration. The term \(A (2\bar{n} + 1)\) captures the effects of thermal motion, which are exacerbated by shuttling operations that increase the vibrational energy of the ion chain.

\par We have validated our parameters against hardware data sheets from Quantinuum and IonQ. We also consider a range of gate improvements (1X to 10X) in our experiments to account for future improvements. A 5X improvement in our setup corresponds to \(\approx 10^{-3}\) depolarising error rates per qubit gate, which is comparable to the best-known devices from Quantinuum and IonQ \cite{H2Racetrack, Chen_2024}.

\textbf{Cooling Model:} \label{subsecCoolingModel} Cooling ions before qubit gates decrease physical error rates at the expense of increased execution times. To model the effect of cooling in the WISE wiring method,
the noise model in Table~\ref{tab:operations} is modified, setting the baseline two-qubit gate error to \(2 \times 10^{-3}\) and the one-qubit gate error to \(3 \times 10^{-3}\), while ignoring heating effects by adding an extra 850 \(\mu s\) to the two-qubit gate time \cite{Pino_2021}. 
 
\subsection{Resource Estimation Model} \label{subsecresourcegstimation}

\par The total number of electrodes \(N_e\) for a trap capacity \(k\), number of junctions \(N_j\), and number of traps \(N_t\) is given by: \( N_e = N_{de} + N_{se} = N_{de/lz} \times N_{lz} + N_{de/jz} \times N_{jz}  + N_{se/z} \times (N_{lz} + N_{jz})
\) where:  \begin{itemize}
    \item The number of linear zones: \(N_{lz} = N_t \times k\),
    \item The number of junction zones: \(N_{jz} = N_j\),
    \item The number of dynamic/shim electrodes per zone: \(N_{lz/de} = 10\), \(N_{jz/de} = 20\), and \(N_{se/z} = 10\) \cite{1000qubits}.
\end{itemize}

\noindent Decreasing the trap capacity increases the number of electrodes for a fixed qubit count. This is because the ratio of junction zones to linear zones, \(N_{jz} / N_{lz}\) increases, so lower trap capacities require more electrodes (since junction zones require more electrodes than linear zones) \cite{1000qubits}.  

\par The controller-to-QPU data rate (Figure~\ref{figQCCDHardwareDesign}) and power dissipation required are calculated using the number of electrodes for the standard QCCD architecture. The data rate between the QPU and its controller is \(
 \approx 50\,\text{Mbit/s} \times N_e
\) while the corresponding power dissipation is \(
\approx 30\,\text{mW} \times N_e
\), where \(N_e\) denotes the number of electrodes. 

\par In the WISE architecture, the data rate is \(
 \approx 50\,\text{Mbit/s} \times N_{\text{DACs}}
\), where the number of DACs is \(
N_{\text{DACs}} \approx 100 + \frac{N_{se}}{100}
\), while \(N_{se}\) denotes the number of shim electrodes. As a result, the WISE architecture scales two orders of magnitude more favourably in terms of data rate compared to the standard architecture, significantly reducing the burden on control electronics \cite{1000qubits}.

\section{Experimental Setup} \label{secExperimentalSetup}

Our experiments benchmark the performance of different combinations of QEC codes and QCCD configurations to answer the architectural questions posed in (§\ref{sec:hardware_tradeoffs}).

\subsection{QEC benchmarks} 
We use three benchmarks for our compiler: 1) repetition code and 2) unrotated surface code are two simple QEC schemes that serve only as baselines for compiler validation, while 3) rotated surface code (Figure~\ref{fig:RotatedSurafecCode4WithQC}) is a more efficient QEC scheme that serves as the primary workload for architectural experiments. We consider code distances $d$ in the range 2 to 20. With increasing code distance, the surface code exponentially reduces errors, but uses a quadratically higher number of qubits (scaling as $2d^2-1$ physical qubits per logical qubit) and communication requirements. Our simulations consider the operation of logical identity in the surface code (essentially \(d\) rounds of parity-check measurements). This operation is selected because maintaining a logical qubit with an error rate lower than the physical error rate during idling is one of the most challenging aspects of quantum error correction. Other logical operations, implemented using transversal gates or lattice surgery, also rely on rounds of parity-checking, so the logical identity serves as a representative workload.

\subsection{Architecture configurations} We explore trap capacities, ranging from 2 to 30, along with the grid, switch and linear connectivities. We also explore the standard choice for control system wiring, where each DAC is connected to one electrode, and the WISE architecture \cite{1000qubits}. Since our study aims to understand the design choices for future systems with potentially improved hardware, we scale the physical error rates by a factor called \emph{gate improvement}. For example, a 10X gate improvement corresponds to every gate having a 10X lower physical error rate and the dephasing physical error rate on idling qubits being 10X less. The gate improvement in our experiment varies from 1X to 10X.

We compile the parity-check circuit for each surface code distance $d$ and architecture configuration combination. Then, we determine architectural and hardware parameters using models from the previous section and use Stim simulations to assess the logical error rate.

\subsection{Metrics}
\par \textbf{Elapsed / QEC Round Time:} The elapsed time is the \textbf{time required to run one round of surface code parity checks when considering gate times and communication times}. Lower elapsed times are better. Prolonged rounds of parity-checking can exacerbate the effects of idling noise, becoming a bottleneck for error correction. Since every logical operation in a fault-tolerant algorithm contains \(d\) rounds of parity-checking to avoid the propagation of errors, the round time directly influences the logical clock speed.

\par \textbf{Logical Error Rate:} The logical error rate quantifies the primary objective of QEC: suppressing quantum errors to levels that enable fault-tolerant computation. The experiment looks to identify configurations capable of achieving a $10^{-9}$ logical error rate, which is a minimum requirement for large-scale algorithms \cite{suppressing_quantum_errors_by_scaling_surface_code}. 

\par \textbf{Number of Movement / Routing Operations:} The number of primitive ion reconfigurations, including split, move, merge, junction entry, exit (t7-t11), plus the number of gate swaps (with each gate swap being 3 two-qubit MS gates (§\ref{subsecTopoCodesBackground})).

\par \textbf{Theoretical Minimum Elapsed Time:} To verify our compiler's performance, we manually compute the best possible elapsed time for specific QEC code and QCCD device combinations. For example, with a trap capacity of 2 a repetition code's structure can be exactly mapped to QCCD. However, since this metric is based on intuitive QEC-device mappings, there may be slight suboptimality in some cases.

\par \textbf{Data Rate and Power:} The data rate is the controller-to-QPU bandwidth required per logical qubit in GBit/s, and the power is the rate of energy dissipation of the QPU per logical qubit, calculated using the resource model in (§\ref{subsecresourcegstimation}).

\subsection{Logical Error Rate Calculation Using Stim} \label{secTestInfrastructure}

\par The logical error rate calculation is performed by interfacing the physical noise model and the execution schedule of the compiled circuit into a noisy quantum circuit in Stim \cite{gidney2021stim}. We use Stim version 1.13.0.

\subsection{Baselines} 
Our QEC compiler (implemented in Python 3.11) is benchmarked against two other trapped-ion QCCD compilers: QCCDSim \cite{murali2020architectingnoisyintermediatescaletrapped} and Muzzle The Shuttle \cite{muzzletheshuttle} in terms of ion movement time and number of movement operations.

\section{Results}\label{sec:results}
\subsection{Accurate and Scalable QEC Compiler}\label{subsecTestAndBenchmark}    

\begin{table}[!htbp]
\centering
\includegraphics[width=238pt]{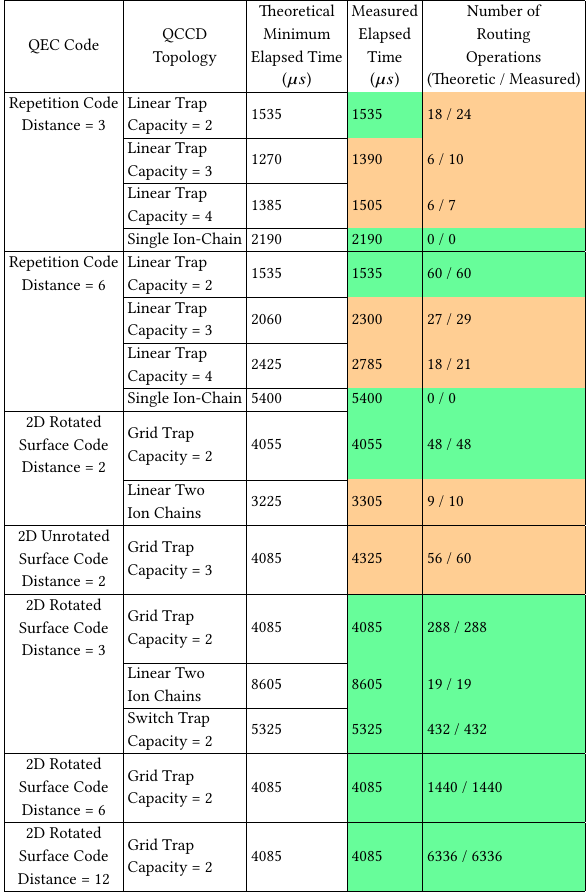}

\caption{Comparison of our QEC compiler against hand-optimised compilation. Our compiler is near-optimal regarding elapsed time and number of routing operations.}
\label{tableIntegrationTest}
\end{table}
\par To validate our compiler, we compare it against hand-optimised baselines. These baselines serve as a correctness check to ensure the compiler recovers known good solutions. Table~\ref{tableIntegrationTest} compares the elapsed time for different QEC code and QCCD device model pairs with the theoretical minimum elapsed time. In 10 out of 16 cases, our compiler matches the expert baseline; in the remaining cases, it is away from the optimum by an average of 1.09X, worst case 1.11X. 
In addition, we test the routing tool in isolation by comparing the theoretical optimal number of routing operations in a schedule to the measured number of routing operations. On average, our compiler is within 1.04X of the theoretical minimum.  While manual mappings work well for small or structurally trivial configurations, it becomes intractable for larger design space explorations.

\begin{table}[!htbp]
\centering
\includegraphics[width=238pt]{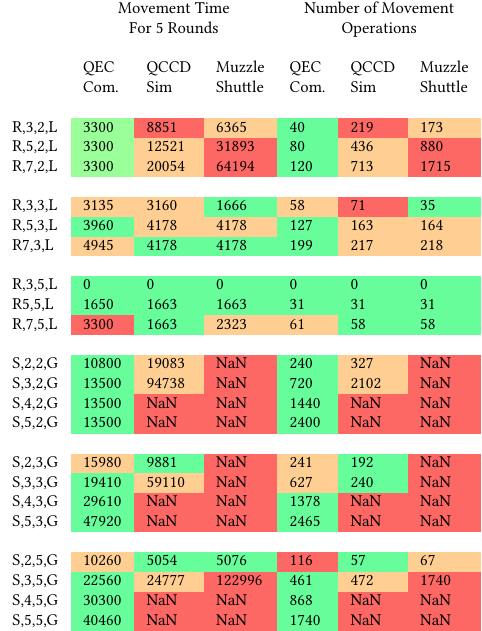}

\caption{Benchmark test of our compiler outlined with other QCCD compilers, namely QCCDSim and MuzzleTheShuttle. Each test determines the movement time and number of movement operations in the compiled schedules for a particular software-hardware configuration. A 4-tuple specifies each configuration: QEC code (R = repetition code, S = 2D Rotated Surface Code), Code Distance, Trap Capacity, and QCCD Communication Topology (L = linear, G = grid). In some cases, a QCCD constraint (§\ref{subsec:ion_routing_algo}) was violated, or the compilation failed, in which cases' NaN' is reported. For each test (row), the compilers are shaded green (best), amber or red (worst). }
\label{tableBenchmarking}
\end{table}

\par Table \ref{tableBenchmarking} compares the performance our QEC compiler QCCDSim \cite{murali2020architectingnoisyintermediatescaletrapped} and Muzzle The Shuttle \cite{muzzletheshuttle}. We benchmarked five rounds of error correction to account for any changes in qubit layout across rounds. For all baselines, the time required to execute gates is the same. Therefore, we focus on movement time (time required for ion reconfigurations) and the number of movement operations. Our QEC compiler achieves an average 3.85X reduction in movement time and an average 1.91X reduction in movement operations compared to the best of the two compilers in each test case. In the best case, the improvement is up to 6.03X. For the 2D rotated surface code, the QEC compiler successfully compiles five rounds of error correction across a wide range of trap capacities and code distances. In contrast, QCCDSim and MuzzleTheShuttle either produce suboptimal schedules or fail to compile entirely, especially at higher code distances. These results show that our compiler is well-suited for architectural evaluations.

\subsection{Choice of Communication Topology}

\begin{figure}[t]
    \centering
    \includegraphics[width=238pt]{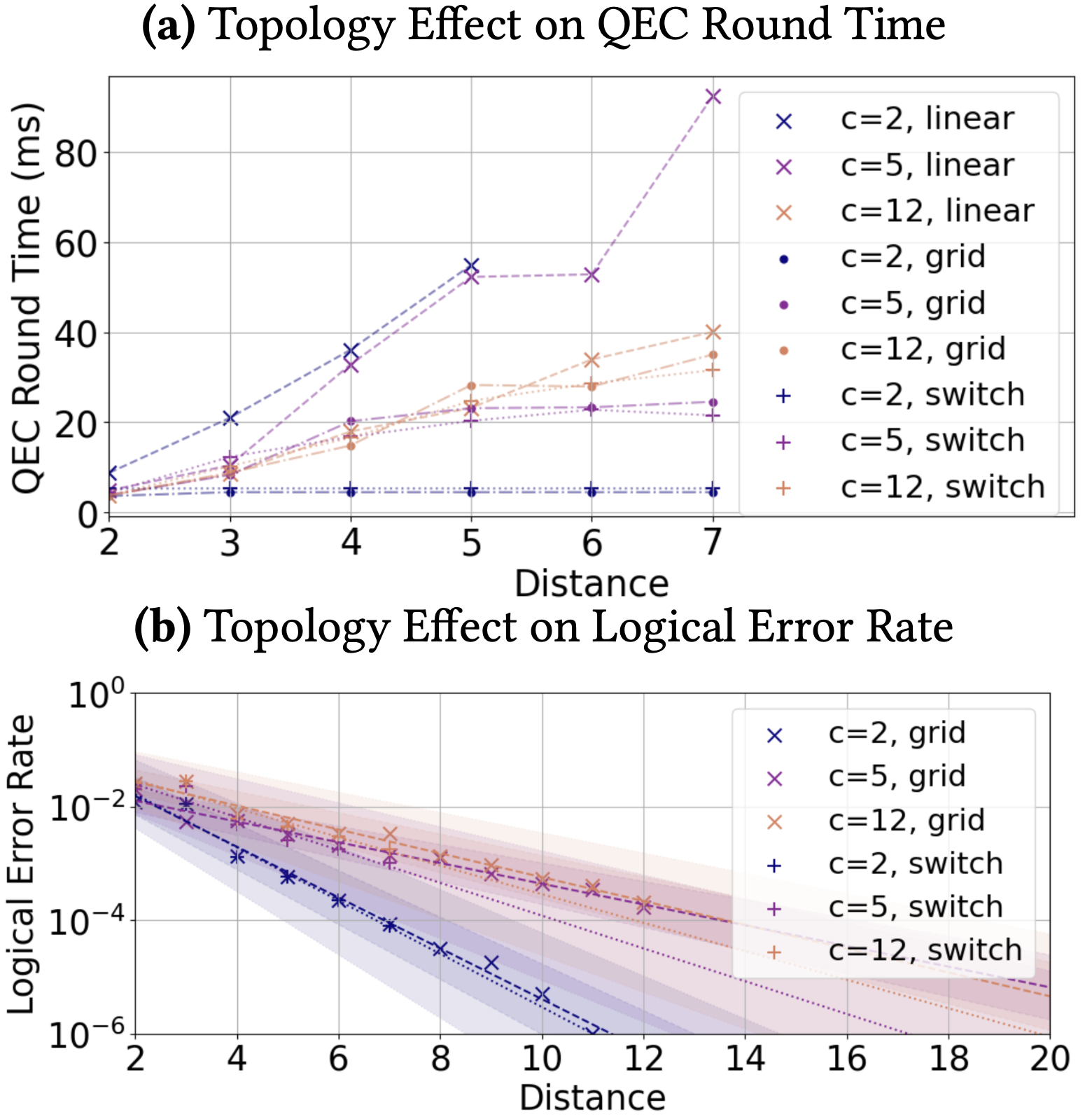}
    
    \caption{%
        \textbf{(a)} Elapsed time per QEC round (y-axis) as a function of code distance (x-axis) for trap capacities 2, 5, and 12, under linear, grid, and all-to-all switch communication topologies. 
        \textbf{(b)} Logical Error Rate as a function of code distance for trap capacities 2, 5, and 12 under the grid and all-to-all switch.
    }
    \label{figElapsedTimeVsTopo}
\end{figure}

\par Figure~\ref{figElapsedTimeVsTopo}(a) compares QEC round time as a function of code distance for the linear, grid, and all-to-all switch communication topologies. We show the results for capacities of 2, 5 and 12, but the trends are similar for other capacities. We make three observations. First, the linear topology exhibits high elapsed times across capacities due to routing congestion. For instance, \(d=5, C=2\) requires over \(\approx 275\)ms per logical identity operation for the linear topology, which is \(\approx 12\)x greater than the switch and grid topologies. This is expected since a linear topology does not match the surface code's requirements. Second, the switch and grid topologies have approximately the same elapsed time. While this is expected for minimal trap capacity where the grid closely matches the surface code's needs, we may expect a switch topology to have a significant advantage for large capacities. This is not the case because operations within a trap get serialised, making it difficult to use the rich connectivity at high trap capacity. Third, only a trap capacity of two with grid or switch topology offers a constant elapsed time, independent of code distance. We discuss this aspect in the following subsection. 

\par Figure~\ref{figElapsedTimeVsTopo}(b) compares the logical error rate versus the code distance and the trap capacity for the grid and switch topologies. Although theoretically, the switch should outperform the grid due to lower contention across routing paths, the difference in logical error rate between the grid and the switch is minor and statistically inconclusive (overlapping error bars). 

\textbf{Our work validates that across trap capacities, the grid topology matches very closely the all-to-all switch both in terms of QEC round time and logical error rate, making it an ideal choice for hardware implementation.} In the following experiments, we use the grid topology.

\subsection{Choice of Trap Capacity}
\label{secEffectOfTrapSize}

\begin{figure}[t]
    \centering
    \includegraphics[width=238pt]{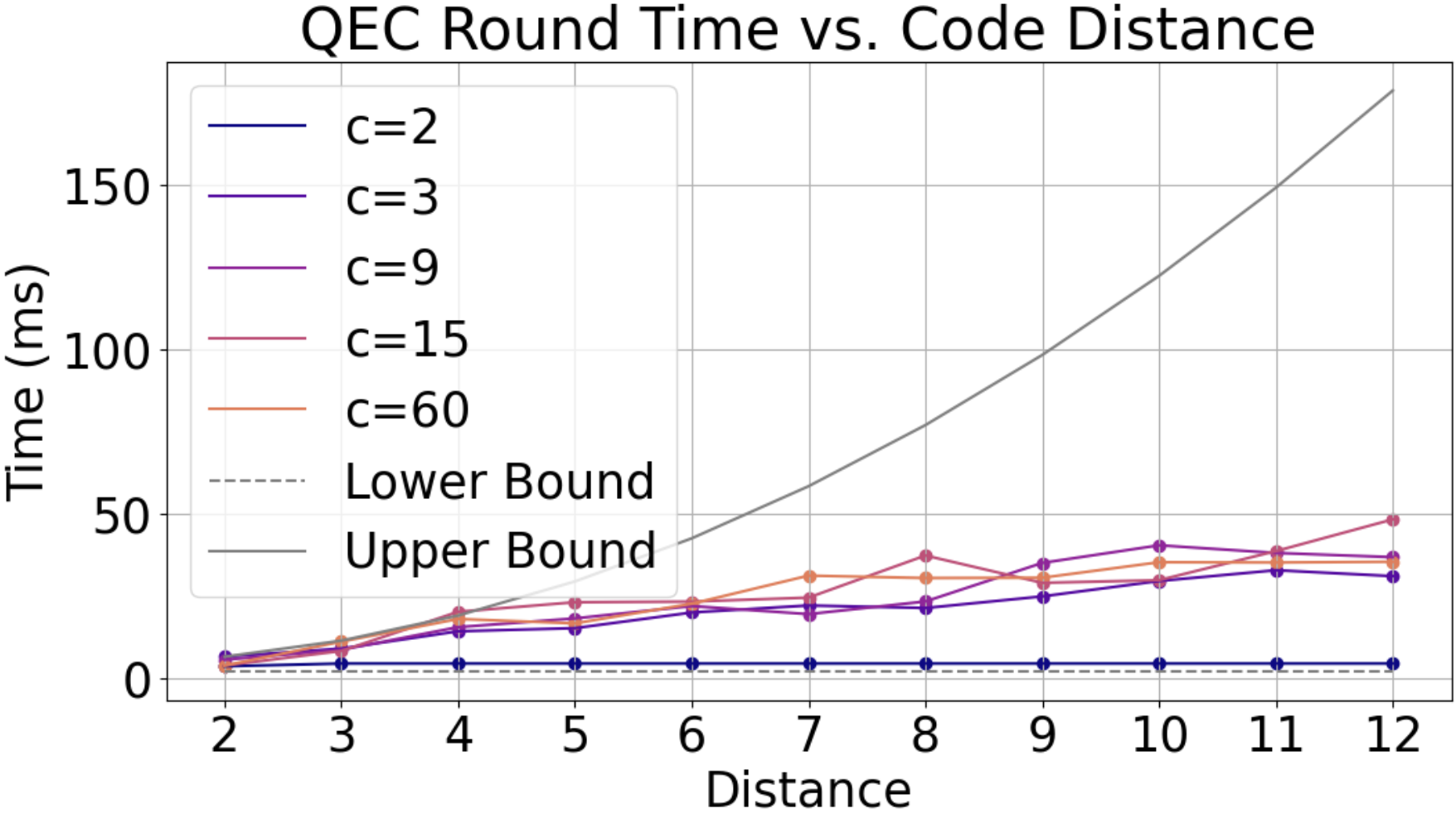}
    \caption{QEC shot time (y-axis) as a function of trap capacity (marked by the legend) and code distance (x-axis). The lower bound (grey dotted) corresponds to the minimal time required (\( 2.5ms \)) for a single round of surface code parity-check operations when there are no ion reconfigurations, and there is complete parallelism. The upper bound represents the elapsed time when all ions are in the same trap, causing complete serialisation. }
    \label{figElapsedTimeWithCapacity}
\end{figure}
\begin{figure}[t]
    \centering
    \includegraphics[width=238pt]{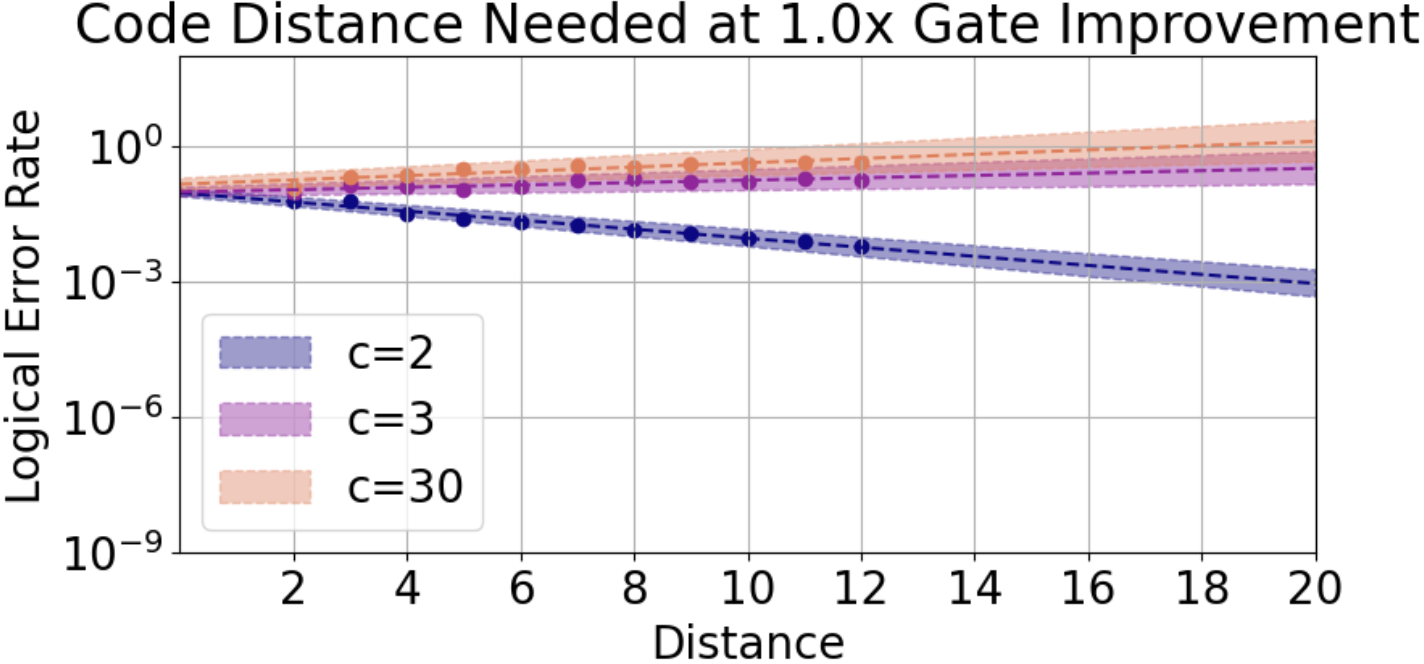}
    \\[6pt]
    \includegraphics[width=238pt]{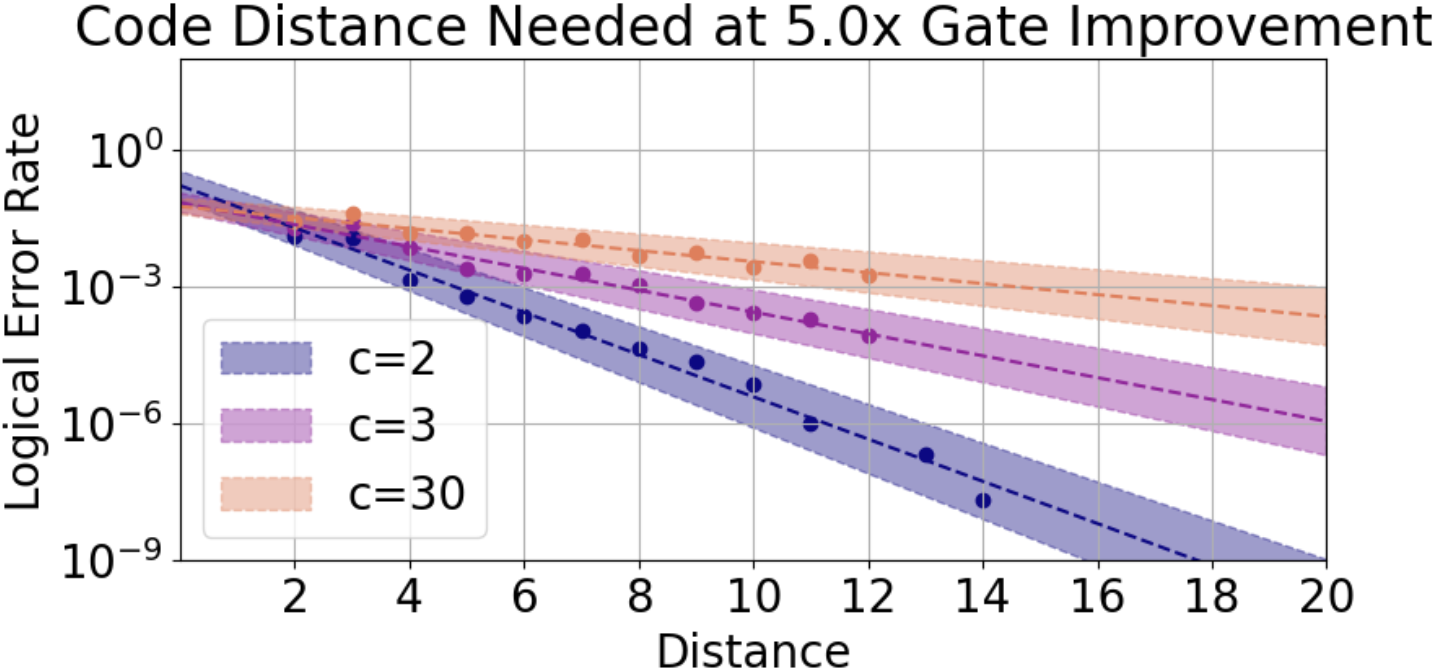}
    \\[6pt]
    \includegraphics[width=238pt]{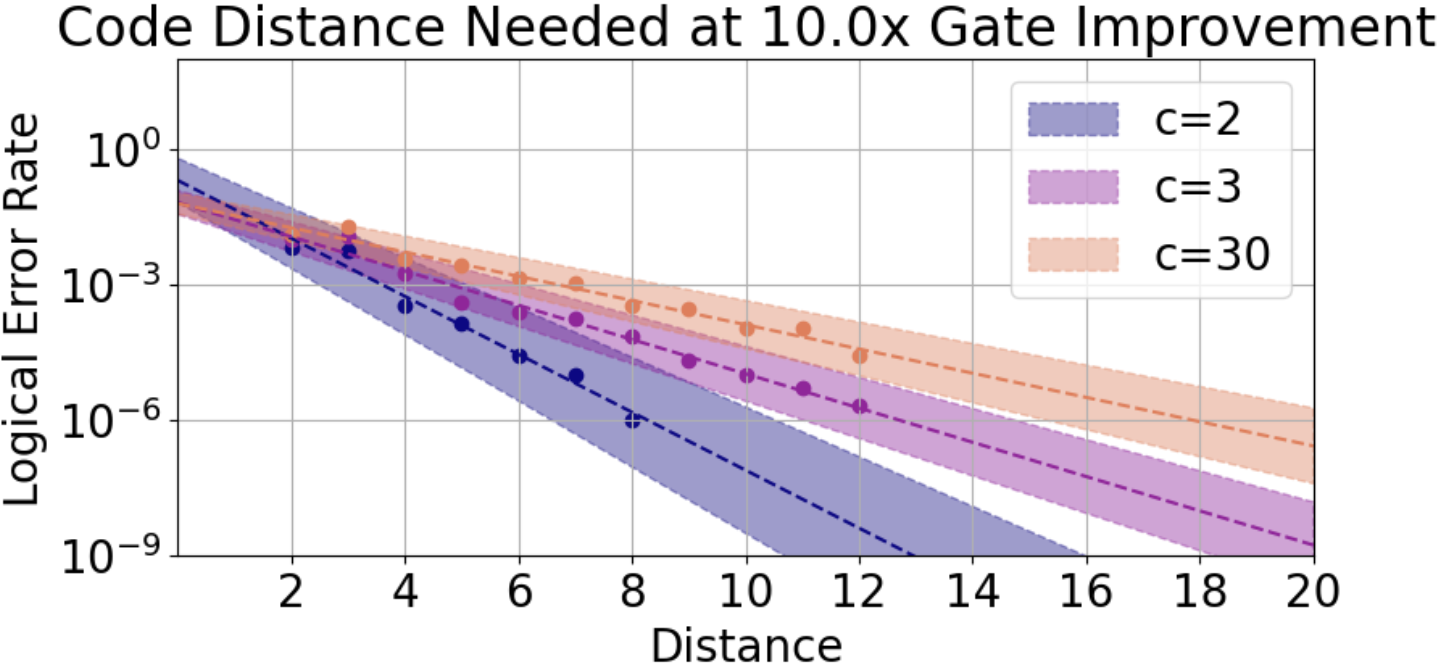}
    \caption{Projections of logical error rate versus code distance for the surface code on a QCCD grid topology at different levels of gate improvement. The target logical error rate of $10^{-9}$ is used to assess practical feasibility, with the x-axis intercept indicating the code distance required to achieve this target. The three axes show projections for 1X, 5X and 10X gate improvements, respectively.}
    \label{figProjectionsGateImprov}
\end{figure}
\textbf{Impact on elapsed time:} Figure \ref{figElapsedTimeWithCapacity} shows the elapsed time for different trap capacities and code distances. A trap capacity of two offers lower elapsed times than higher capacities. These elapsed times are also close to the theoretical lower bound. This is surprising because a capacity of two incurs the maximum number of communication operations; a larger trap capacity reduces the need for reconfiguring ions, as ancilla qubits are more likely to be located with their data qubits. However, using a capacity of two maximises the number of gates that can be executed in parallel; a larger capacity serialises more operations within a trap. Our work shows that maximising parallelism is more important for efficiently mapping surface codes onto QCCD systems and offering the best runtimes for large-scale applications that may use millions of QEC rounds. 

Further, a trap capacity of two also offers constant cycle time irrespective of code distance, whereas higher capacities see cycle times grow with code distance. Although this was not a design goal, constant cycle time is an elegant architectural design point that mirrors the fixed cycle time of classical processors. Having this parameter independent of the error correction parameters and application demands will benefit abstraction and predictable system performance in the long term. Importantly, a trap capacity of two does not trade performance for consistency; it also achieves the lowest logical error rates (Figure~\ref{figProjectionsGateImprov}).

\textbf{Impact on logical error rate:} Figure~\ref{figProjectionsGateImprov} evaluates the effect of trap capacity on the logical error rate of the surface code. We use three physical gate improvement scenarios, with 1X corresponding to pessimistic scaling of current systems, 5X corresponding to optimistic scaling of current systems, and 10X corresponding to a future improved system. Across gate improvement scenarios, a trap capacity of two outperforms higher capacities by one to two orders of magnitude in logical error rate. This is because a parallel system with very small traps can better localise error propagation and keep gate error rates well below the code threshold (§\ref{def:code_threshold}), enabling the exponential logical error rate suppression. Even with future improvements in physical gates, a trap capacity of two remains an excellent choice for logical qubit design on QCCD systems.

Further, early scientific applications are expected to require at least a logical error rate of $10^{-9}$ to offer advantages over classical computing. From Figure \ref{figProjectionsGateImprov}, it is clear that to achieve a low logical error rate we can either implement high code distances (increasing the number of physical qubits per logical qubit) or improve the physical gates. Trap capacity of two paired with a code distance of 13, with a 10X improvement in physical gate quality, is a feasible design point for quantum advantage experiments. If a 10X physical improvement proves infeasible in the coming years, increasing the code distance to 18 would offer the same logical qubit quality.

\begin{figure}[!htbp]
    \centering
    \includegraphics[width=238pt]{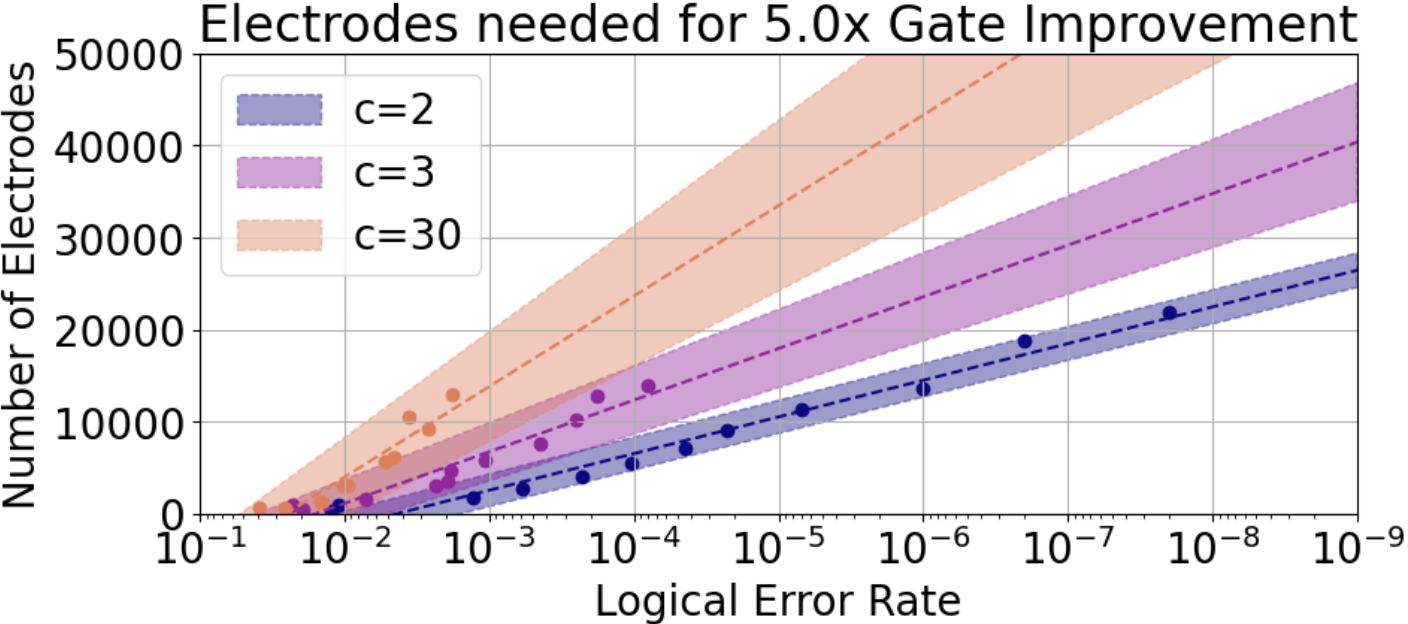}
    \caption{Projected number of electrodes required to achieve a target logical error rate under a 5x gate improvement scenario for different trap capacities.}
    \label{figElectrodesNeeded}
\end{figure}

\par \textbf{Impact on hardware footprint:} Figure~\ref{figElectrodesNeeded} shows the number of electrodes required to implement a QCCD device across different trap capacities. The number of electrodes is an important indicator of the hardware cost~(§\ref{subsecresourcegstimation}). Our results show that all trap capacities are expensive from a hardware perspective under the standard control wiring scheme, but \textbf{trap capacity two is the most hardware-efficient design point}, reducing the electrode counts needed to achieve a given logical error rate by several orders of magnitude compared to higher trap capacities. This is surprising because junctions in a QCCD system require 2X electrodes compared to traps. Therefore, as the trap capacity increases, the number of junctions needed in the design decreases. A design with a higher capacity is expected to offer lower electrode counts when viewed purely from a hardware perspective. However, when viewed from the standpoint of implementing logical qubits, increasing the trap capacity leads to worse logical error rates (Figure~\ref{figProjectionsGateImprov}). In turn, a given logical error rate requirement necessitates the use of logical qubits with higher code distances, which increases the overall physical qubit count and the number of junctions and traps and, therefore, requires large electrode counts.

\textbf{Unlike prior NISQ studies, which recommend the use of traps with capacity in the range of 20-30 ions \cite{murali2020architectingnoisyintermediatescaletrapped}, we advocate the use of a trap capacity of two to obtain logical qubits with hardware efficiency, low error rates, and a constant runtime regardless of code distance.} 

\begin{figure}[!htbp]
    \centering
    \includegraphics[width=238pt]{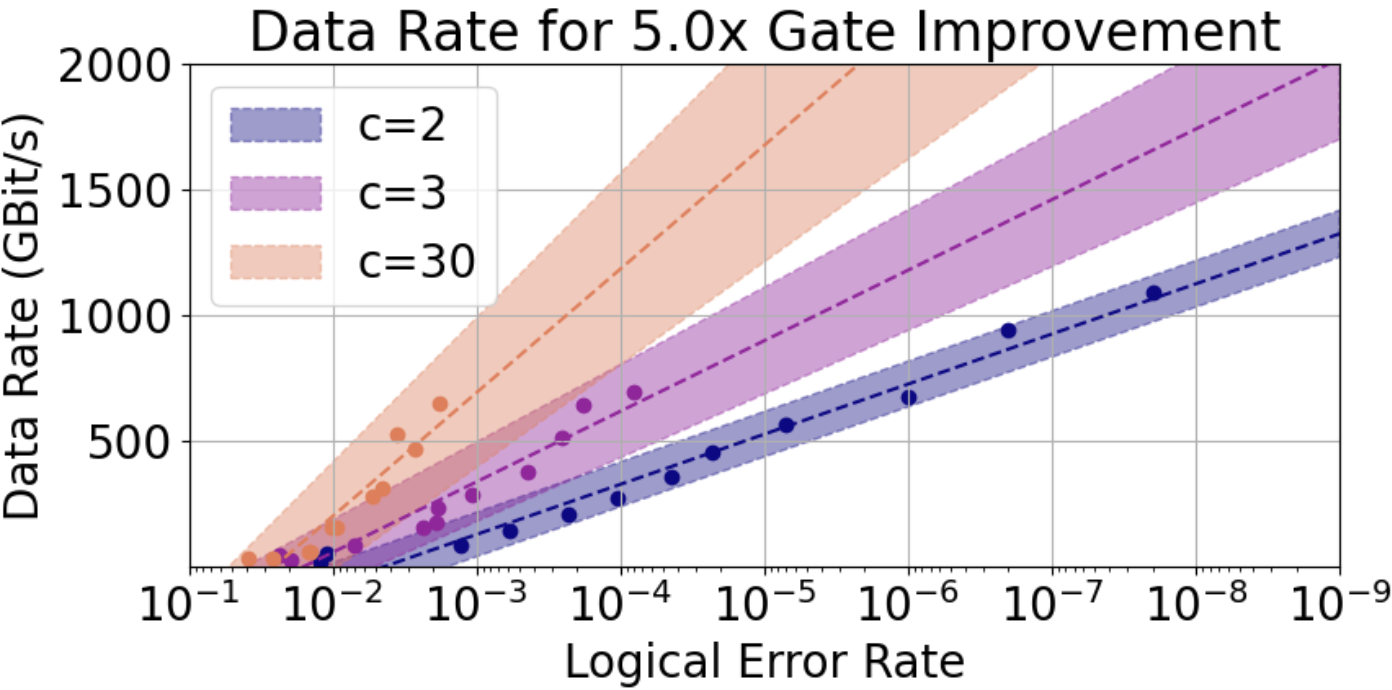}
    \caption{Hardware requirements for achieving a target logical error rate under a 5x gate improvement scenario across different trap capacities (\(c\)). The axis shows the required data rate between the QPU and the controller. A trap capacity of \(c=2\) minimises both power dissipation and data rate demands at a logical error rate of \(10^{-9}\). However, even in this optimal case, achieving \(10^{-9}\) necessitates an impractical \(1.3\) Tbit/s communication link and  \(\approx 780\) W of power dissipation.}
    \label{figHardwareRequirements}
\end{figure}

\subsection{Choice of wiring method} 
\par At a trap capacity of two, with every \(\approx 5,000\) additional electrodes, we obtain an \(\approx 10\)X decrease in logical error rate. Although this represents the best scaling observed, it remains far from practical. Figure \ref{figHardwareRequirements} confirms that the data rate and power requirements for a standard QCCD architecture quickly reach impractical levels as the system scales. In particular, a single logical qubit with an error rate of $10^{-9}$ demands a power consumption of more than 780 Watts. A system with a few thousand logical qubits and much lesser logical error rates is required for practical quantum applications and may lead to trapped-ion systems requiring tens to hundreds of megawatts of power per system.

\begin{figure}[t]
    \centering
    \includegraphics[width=250pt]{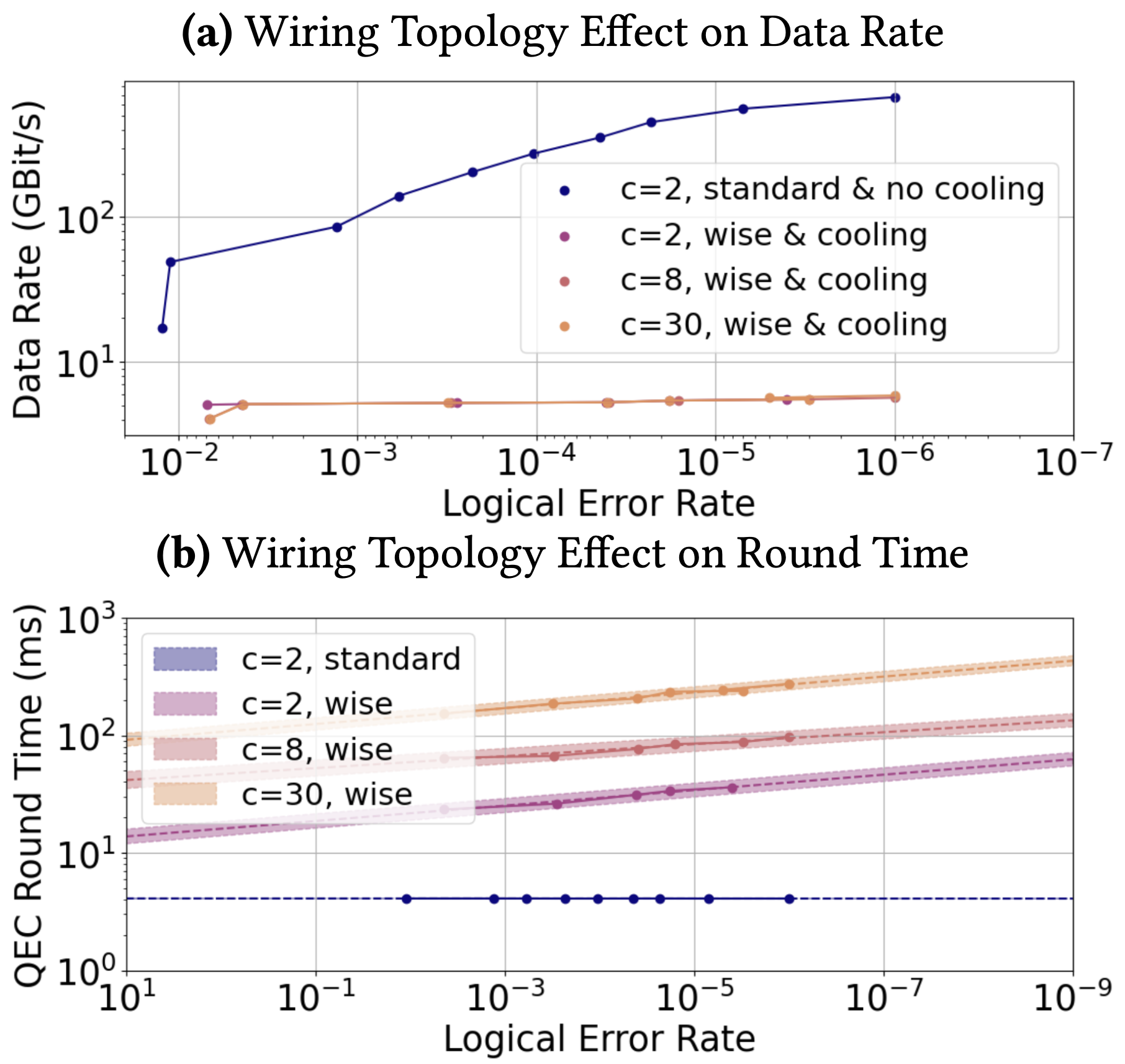}

    \caption{\textbf{(a)} Data rate comparison between the standard architecture without cooling and WISE architecture with cooling, under a 5X gate improvement. Cooling improves data rate scaling across all trap capacities for the WISE architecture, allowing low logical error rates at modest data rate requirements compared to standard capacity‐2 systems. \textbf{(b)} Elapsed QEC shot time versus target logical error rate under a 5X gate improvement. In the WISE architecture with cooling, logical scale quadratically with code distance, leading to a logical clock speed of \(\approx10^{-1}\) operations per second for a \(10^{-9}\) target error rate. In contrast, the standard, no cooling, trap capacity two architecture exhibits linear scaling of cycle times with increasing code distance.}
    \label{figStandardWiseCooling}
\end{figure}

A key power bottleneck in the standard architecture is that each electrode is wired to a separate DAC. WISE \cite{1000qubits} overcomes this with a more intelligent wiring mechanism, trading off execution time for reduced power consumption. \textit{Which mechanism is the most suitable for logical qubit implementation?} Figure \ref{figStandardWiseCooling}(a) compares the data required for WISE and the standard wiring mechanism. For the standard mechanism, we only use trap capacity 2. Whereas, for WISE, we examine trap capacities ranging from 2 to 30 but only show the curves for three capacities, since the trends are similar at other capacities. Compared to the standard architecture, WISE achieves an improvement of more than two orders of magnitude in data rate (and, therefore, in power consumption). 

WISE requires cooling support from the hardware to reduce physical noise (our simulations with no cooling for WISE indicated that it could not scale beyond a logical error rate of $10^{-4}$ without). As a result, contrary to the standard architecture, trap capacity two is not more hardware-efficient than other trap capacities in the WISE architecture. However, smaller traps still achieve the lowest QEC round times while maintaining modest data rate requirements. \textbf{In both control systems, designing traps to be as small as possible remains optimal for surface code implementation.} 

Figure~\ref{figStandardWiseCooling}(b) compares the elapsed time at different logical error rates. For the WISE architecture, the elapsed time scales in proportion to the desired logical error rate. For every 10X improvement desired in the logical error rate, the elapsed time increases by 1.17X. WISE suffers from limited transport flexibility, allowing only one transport operation at a time. Under an odd-even sort global reconfiguration scheme \cite{1000qubits}, this limitation results in logical clock speeds that are up to $25$X longer than those of standard QPUs, for logical error rates near $10^{-9}$. This runtime increase is acceptable for near-term fault-tolerant applications such as quantum dynamics \cite{vandam2024usingazurequantumresource}. However, for large applications such as factoring, which already require month-long computations on trapped ion systems\cite{Lekitsch_2017}, such a runtime increase will lead to impractical executions that run over a year.

Therefore, \textbf{we observe a power vs. cycle time trade-off in current wiring mechanisms for QCCD trapped ion systems.} Multiplexed wiring mechanisms lead to low power but very long execution times, while direct wiring of DACs to electrodes offers low execution times with high power consumption. For scaling trapped ions to the regime of several hundred logical qubits, we need to go beyond existing control system designs. We require novel architectures that offer high-performance executions with low power needs.

\section{Related Work}

\par This work builds on previous advancements in QCCD system architecture and QEC optimisation. For instance, Guti\'errez et al. \cite{transversality_lattice_surgery} inspire the test infrastructure to validate executable QCCD circuits. The relevance of compiler-driven architectural co-design for QCCD systems is demonstrated by Murali et al. \cite{murali2020architectingnoisyintermediatescaletrapped}, which examines the influence of micro-architectural choices on the performance of NISQ algorithms. Similarly, Wu et al. \cite{synthesis_framework_stitching} address the challenges in bridging quantum hardware and QEC codes by proposing a framework for efficient implementation and optimisation of surface codes for superconducting architectures. This study extends these concepts by tailoring a QEC compiler to the specific demands of QCCD-based systems, aiming to provide a systematic approach to co-designing hardware and software for fault-tolerant quantum computing. 

\par While there exist QCCD compilers for QEC other than the two benchmarked in (§\ref{subsecTestAndBenchmark}), such as the MQTIonShuttler \cite{schoenberger2024shuttlingscalabletrappedionquantum}, we do not benchmark our compiler against these, since they assume distinct memory and processing zones in their QCCD architecture, which is not suitable for surface code implementation. TISCC \cite{Leblond_2023} fixes the trap capacity as two and the standard grid topology \cite{Lekitsch_2017}, then compiles and simulates high-level logical circuits into a quantum circuit on physical qubits using the surface code. The compiler does not map to primitive QCCD directly but uses the performance models of these primitives for resource estimation. 

\textbf{Consideration of Limiting Factors:} In contrast to superconducting platforms, decoder runtimes are not the limiting factor for ion-trap systems since their cycle time is considerably longer. Specialist hardware is already available for the fast decoding of surface codes up to a distance of 8 \cite{Barber_2025}. 

\par We recognise that there are other architectural challenges not addressed: integrating many logical qubits in monolithic QCCD systems (since such scaling will require networking between multiple ion-trap systems), general noise inhomogeneity across the ion chain, and universal gate set implementation. However, if lattice surgery is used to perform entanglement between logical qubits, only boundary qubits of the two logical qubits will need to participate in such circuits, leaving the bulk of the surface code intact. Since the quantum circuits from lattice surgery are very similar in structure to the circuits within one surface code qubit, we expect our results to hold.

\section{Conclusion}
TI qubit technology is at the threshold of supporting systems with several logical qubits. Current demonstrations of logical qubits are limited to small systems of less than 60 physical qubits. To scale up to systems with several hundred physical qubits (tens of logical qubits), we need to understand what the right trap capacities and topologies are and how control systems must be designed to support QEC workloads. The TI community has been exploring these choices for several years, with 1) monolithic, large trap capacity devices (e.g., IonQ Forte) 2) QCCD devices with small trap capacities (e.g., Quantinuum H2) 3) architecture research showing the value of QCCD systems with 15-25 ions per trap \cite{murali2020architectingnoisyintermediatescaletrapped} and 4) other manual design efforts \cite{1000qubits, Lekitsch_2017, valentini2024demonstrationtwodimensionalconnectivityscalable}. 

We conduct a systematic architectural design exploration for implementing logical qubits on TI systems.  Unlike prior studies, our work shows the value of using a trap capacity of two to obtain high-performance, hardware-efficient, low error rate logical qubits with a constant runtime irrespective of QEC code distance. Our work also shows the importance of co-designing control architectures with QEC needs. 

To scale TI systems to the sizes required for practical quantum advantage, our architectural guidance and toolflow are likely to be very important. 

\bibliographystyle{ACM-Reference-Format}
\balance
\bibliography{paper}

\end{document}